\documentclass[aps,prc,superscriptaddress,reprint]{revtex4-1}
\usepackage{graphicx} 
\usepackage{epstopdf}
\usepackage{dcolumn} 
\usepackage{bm}
\usepackage{hyperref}
\usepackage{array}
\usepackage{setspace}
\usepackage{amsmath}
\usepackage{booktabs}
\hypersetup{
    colorlinks=true,
    linkcolor=blue,
    filecolor=gray,
    urlcolor=blue,
    citecolor=blue,
}
\begin{document}
\title{Assessing the Impact of Nuclear Mass Models on the Prediction of Synthesis Cross Sections for Superheavy Elements}
\author{Chang Geng}
\affiliation{College of Physics Science and Technology, Yangzhou University, Yangzhou 225009, China}
\author{Peng-Hui Chen}
\email{Corresponding author: chenpenghui@yzu.edu.cn}
\affiliation{College of Physics Science and Technology, Yangzhou University, Yangzhou 225009, China}
\author{Fei Niu}
\affiliation{Henan Police College, Zhengzhou 450046, China}

\author{Zu-Xing Yang}
\affiliation{RIKEN Nishina Center, Wako, Saitama 351-0198, Japan}

\author{Xiang-Hua Zeng}
\affiliation{College of Physics Science and Technology, Yangzhou University, Yangzhou 225009, China}

\author{Zhao-Qing Feng}
\affiliation{School of Physics and Optoelectronics, South China University of Technology, Guangzhou 510641, China}

\date{\today}
\begin{abstract}

Within the framework of the dinuclear system model, this study delves into the impact of various nuclear mass models on evaluating the fusion probability of superheavy nuclei. Nuclear mass models, as crucial inputs to the DNS model, exhibit slight variations in binding energy, quadrupole deformation, and extrapolation ability; these subtle differences can significantly influence the model's outcomes. Specifically, the study finds that nuclear mass plays a pivotal role in determining fusion probability, and Q-value.
By numerically solving a set of master equations, the study examines how binding energies from different mass models affect the fusion probability of colliding nuclei, taking the example of $^{48}$Ca + $^{243}$Am $\rightarrow$ $^{291}$Mc. A careful analysis of the potential energy surface (PES) reveals that the inner fusion barriers lead to variations in fusion probabilities.
Importantly, the study demonstrates that the synthesis cross sections of superheavy nuclei calculated using different nuclear mass models align well with experimental data, falling within an error range of one order of magnitude. This finding underscores the reliability of our model predictions.
Looking ahead, the study utilizes five distinct nuclear mass models to predict the synthesis cross sections of superheavy elements 119 and 120, along with their associated uncertainties. These predictions offer valuable insights into the feasibility of synthesizing these elusive elements and pave the way for future experimental explorations.

\begin{description}
\item[PACS number(s)]
25.70.Jj, 24.10.-i, 25.60.Pj
\end{description}
\end{abstract}
\maketitle
\section{Introduction}

The exploration of synthesis superheavy nuclei and their decay characteristics represents a pivotal endeavor in natural sciences, occupying a cutting-edge position within nuclear physics research. These nuclei not only hold immense significance in probing the boundaries of the atomic mass number and testing models of atomic nucleus shells but also emerge as an ideal medium for studying the behavior of nuclear matter under intense Coulombic conditions.
Since the 1960s, the experimental realization of superheavy nuclei has been a subject of profound interest. Prominent theoretical physicists like Myers and Nilsson had envisioned the existence of "islands of stability" for superheavy nuclei\cite{THOENNESSEN2013312}, pinpointing Z = 114 and N = 184 based on the nuclear shell model. The endeavor to synthesize these new heavy and superheavy nuclei is paramount in expanding our nuclide map and reaching these elusive "islands of stability"\cite{PhysRevLett.132.072501,PhysRevLett.132.072502}. This pursuit promises to unlock deeper insights into the fundamental nature of nuclear matter.

In recent decades, significant progress has been achieved in the experimental synthesis of new superheavy elements (SHEs) with atomic numbers ranging from Z=104 to 118. The foremost laboratories leading this endeavor are the Heavy Ion Acceleration System in Germany (GSI), the Dubna Radioactive Ion Beam Accelerator in Russia, the RIKEN Accelerator Research Facility in Japan, the Livermore and Berkeley National Labs in the USA, and IMP-HIRFL in China. These laboratories have successfully filled the seventh period of the periodic table by synthesizing SHEs primarily through fusion-evaporation reactions.
Fusion-evaporation reactions can be categorized as hot or cold, depending on the excitation energy of the superheavy compound nuclei. Hot fusion reactions involve excitation energies exceeding 30 MeV, while cold fusion reactions have excitation energies between 0 and 30 MeV. Utilizing hot fusion reactions, elements such as Rf\cite{HEINLEIN1978407}, Db\cite{1970flog,PhysRevLett.24.1498}, Sg\cite{PhysRevLett.33.1490}, Fl\cite{PhysRevLett.105.182701}, Mc\cite{PhysRevC.69.021601}, Lv\cite{PhysRevC.69.054607}, Ts\cite{PhysRevLett.104.142502}, and Og\cite{PhysRevC.74.044602} have been experimentally realized. On the other hand, cold fusion reactions have led to the synthesis of Sg\cite{1975Og}, Bh\cite{1981ZPhyA300107M}, Hs\cite{article84mu}, Mt\cite{articlehof95}, Ds\cite{articlehof95}, Rg\cite{articlehof111}, Cn\cite{Hof96112}, and
Nh\cite{doi:10.1143/JPSJ.73.2593}.
Notably, the Chinese group SHE has contributed to this field by synthesizing superheavy isotopes such as $^{259}$Db\cite{epjagan01}, $^{265}$Bh\cite{06-4-8npr06} and $^{271}$Ds\cite{cpl12zhang} at the Institute of Modern Physics (IMP, Lanzhou). Additionally, there has been considerable research into exotic actinide nuclei, indicating China's growing foothold in superheavy nucleus research. Recently, the Institute has enhanced its capabilities by modifying the High-Density Superconducting Linear Collider for superheavy nuclei research and has made promising initial progress.
However, the experimental synthesis of superheavy nuclei is a challenging, time-consuming, and costly process, often yielding extremely small cross sections. This is especially true for newer SHEs with atomic numbers Z=119 and 120. Therefore, reliable theoretical studies that can guide experiments are crucial. In recent years, both theorists and experimentalists have shown increasing interest in synthesizing SHEs using near-barrier heavy-ion collisions, including fusion-evaporation and multinucleon reactions. These low-energy heavy-ion collisions involve complex dynamic processes with multiple time-evolving degrees of freedom, such as radial distance, charge and mass asymmetry, quadrupole deformation, and incident energy\cite{nst2021niu}.

Nuclear theorists have long been intrigued by the intricate mechanism of synthesizing superheavy nuclei. In pursuit of this elusive goal, numerous theoretical models have emerged to decipher the optimal conditions for their formation. These models aim to predict with precision the most favorable projectile-target combinations, the ideal incident energies, the preferred evaporation channels, and the maximum achievable synthesis cross sections. The quest to synthesize these exotic nuclei is not only driven by the desire to expand the known limits of nuclear existence but also to unravel the exotic properties that they are believed to exhibit.
To date, a plethora of models have been developed to investigate the synthesis of heavy and superheavy nuclei, each offering a unique perspective on the underlying physics. Prominent among these are the time-dependent Hartree-Fock (TDHF) approach\cite{GUO2018401,10.3389/fphy.2019.00020,MARUHN20142195}, the improved quantum molecular dynamics (ImQMD) model\cite{PhysRevC.65.064608,PhysRevC.88.044611,PhysRevC.88.044605}, the GRAZING model based on dynamic analysis\cite{Grazing}, and the dinuclear system (DNS) model\cite{FENG200650,PhysRevC.91.011603,PhysRevC.89.024615,epja20gga,07fengcpl}. While each model brings its own set of strengths to the table, they often diverge in their predictions due to differences in their treatment of the complex dynamical processes involved.
In this study, we seek to explore the impact of nuclear mass models on the prediction of synthesis cross sections for superheavy nuclei within the framework of the DNS model. Specifically, we consider five distinct nuclear mass models: the Finite-Range-Droplet-Model (FRDM12)\cite{MOLLER20161}, the Koura-Tachibana-Uno-Yamada (KTUY05)\cite{10.1143/PTP.113.305}, the Weizsacker-Skyrme (WS4)\cite{WANG2014215}, the Little Droplet model (MS96)\cite{MYERS19661}, and the Hartree-Fock-Bogoliubov (HFB02)\cite{ELBASSEM201722}. The DNS model, known for its ability to self-consistently account for key physical quantities such as shell effects, dynamical deformation, fission, quasi-fission, deep inelastic collisions, and odd-even effects\cite{PhysRevC.91.011603,FENG200650,FENG201082,07fengcpl,Chen2023}, provides a robust platform for our investigation.
Through a comprehensive analysis, we aim to elucidate the influence of different nuclear mass models on the predicted synthesis cross sections and to identify the most promising avenues for the experimental realization of superheavy nuclei. Our findings not only contribute to the fundamental understanding of nuclear synthesis but also have implications for future experiments in this exciting frontier of nuclear physics.

In this study, we employ the DNS model to calculate the reaction $^{48}$Ca + $ ^{243}$Am $\rightarrow$ $^{291}$Mc$^*$, considering five atomic nuclei mass tables in the fusion stage separately, and compare them with the experimental data to further investigate their effects. We also systematically calculated four reaction systems 
$^{51}$V + $^{248}$Cm $ \rightarrow $ $^{299}$119 $^*$, $^{54}$Cr + $^{243}$Am $ \rightarrow $ $^{297}$119$^*$, $^{54}$Cr + $^{248}$Cm $\rightarrow $ $^{302}$120$^*$,$^{55}$Mn+$^{243}$Am $\rightarrow $ $^{298}$120 $^*$, 
to synthesize elements 119 and 120.
The manuscript is structured as follows: Section \ref{sec2} provides a concise overview of the DNS model, outlining its key features and underlying principles. Section \ref{sec3} presents the calculated results and detailed discussions, comparing the outcomes obtained using different mass tables and exploring the implications for the synthesis of superheavy elements. Finally, Section \ref{sec4} summarizes the main findings and conclusions of our study.

\section{Model description}\label{sec2}

The dinuclear system (DNS) model is utilized to elucidate the mechanisms of near-barrier collisions, depicting them as quasi-molecular configurations wherein two colliding partners retain their individuality throughout the fusion process. This model comprises three distinct stages: capture, fusion, and survival.
During the capture stage, the colliding partners overcome the Coulomb barrier, resulting in the formation of a composite system. Subsequently, in the fusion stage, the kinetic energy and angular momentum are dissipated within the composite system, facilitating nucleon transfer between the touching colliding partners. This transfer continues until all nucleons from the projectile nuclei are transferred to the target nuclei, ultimately leading to the formation of compound nuclei with specific excitation energy and angular momentum.
Finally, in the survival stage, the excited compound nuclei undergo de-excitation through the emission of light particles, ultimately resulting in the ground state nuclei.

Within the DNS model, the evaporation residual cross sections of the superheavy nuclei are written as
\begin{eqnarray}
\sigma_{\mathrm{ER}}\left(E_{\mathrm{c} . \mathrm{m} .}\right)  =  \frac{\pi \hbar^{2}}{2 \mu E_{\mathrm{c} . \mathrm{m} .}} \sum_{J  =  0}^{J_{\max }}(2 J+1) T\left(E_{\mathrm{c} . \mathrm{m} .}, J\right) \nonumber \\
\times P_{\mathrm{CN}}\left(E_{\mathrm{c} . \mathrm{m} .}, J\right) W_{\mathrm{sur}}\left(E_{\mathrm{c} . \mathrm{m} .}, J\right).
\end{eqnarray}
Here, the penetration probability $T(E_{\rm c.m.},J)$ is calculated by the empirical coupling channel model\cite{FENG200650}. The fusion probability $P_{\rm CN}(E_{\rm c.m.},J)$ is the formation probability of compound nuclei\cite{PhysRevC.80.057601,PhysRevC.76.044606}. The survival probability $W_{\rm sur}$ is the probability of the highly exciting compound nuclei surviving by evaporating light particles.

\subsection{ Capture probability}

The capture cross section is written as
\begin{eqnarray}
\sigma_{\text {cap }}\left(E_{\mathrm{c} . \mathrm{m} .}\right)=\frac{\pi \hbar^{2}}{2 \mu E_{\mathrm{c} . \mathrm{m} .}} \sum_{J}^{J_{\rm max}} (2 J+1) T\left(E_{\mathrm{c} . \mathrm{m} .}, J\right).
\end{eqnarray}
Here, the $T(E_{\rm c.m.,J})$ is penetration probability evaluated by the Hill-Wheeler formula \cite{PhysRev.89.1102} within the barrier distribution function.
\begin{eqnarray}\label{hwl}
&T(E_{\mathrm{c.m.}},J)=\int \frac{f(B)}{1+\exp\Big\{-\frac{2\pi}{\hbar\omega(J)}[E_{\mathrm{c.m.}}-B-I ]\Big\}}\mathrm{d}B.
\end{eqnarray}
Here $\hbar \omega (J)$ is the width of the parabolic barrier at $R_{\rm B}(J)$. The normalization constant is $\int f(B)dB=1$. The barrier distribution function is the asymmetric Gaussian form \cite{FENG200650,PhysRevC.65.014607}

The interaction potential of two colliding partners is written as 
\begin{gather}
 V_{\rm CN}(r,\beta_{1},\beta_{2},\theta _{1},\theta_{2})=V_{\rm C}(r,\beta_{1},\beta_{2},\theta _{1},\theta_{2}) + \nonumber\\
V_{N}(r,\beta_{1},\beta_{2},\theta _{1},\theta_{2})
+\frac{1}{2} C_{1}(\beta_{1}-\beta_{1}^{0} )^{2}
+\frac{1}{2} C_{2}(\beta_{2}-\beta_{2}^{0} )^{2}.   
\end{gather}

The 1 and 2 stand for the projectile and the target. The $R=R_1+R_2+s$ and s are the distance between the center and surface of the projectile-target. The $R_1$, $R_2$ are the radii of the projectile and target, respectively. The $\beta_{1(2)}^0$ are the static quadrupole deformation. The $\beta_{1(2)}$ are the adjustable quadrupole deformation\cite{Cheng2022}. 
The nucleus-nucleus potential is calculated by the double-folding method \cite{PhysRevC.80.057601,PhysRevC.76.044606, J.Mod.Phys.E5191(1996)}.
The Coulomb potential is evaluated by Wong's formula \cite{PhysRevLett.31.766}.

\subsection{Fusion probability}

 In the fusion process, the probability of the fragments is evaluated by numerically solving a set of master equations. The term of probability $P(Z_1,N_1,E_1,t)$ contain proton number, neutron number of $Z_1,$ and $N_1$, and the internal excitation energy of $E_1$. The master equation is written as \cite{PhysRevC.76.044606,FENG201082,FENG200933}
\begin{eqnarray}
&& \frac{d P(Z_1,N_1,E_1,t)}{d t} =  \nonumber \\ && \sum \limits_{Z'_1}W_{Z_1,N_1;Z'_1,N_1}(t) \Big[ d_{Z_1,N_1}P(Z'_1,N_1,E'_1,t) \nonumber \\ && - d_{Z'_1,N_1}P(Z_1,N_1,E_1,t) \Big] + \nonumber \\ &&
 \sum \limits_{N'_1}W_{Z_1,N_1;Z_1,N'_1}(t)\Big[d_{Z_1,N_1}P(Z_1,N'_1,E'_1,t) \nonumber \\ && - d_{Z_1,N'_1}P(Z_1,N_1,E_1,t) \Big] - \nonumber \\
 &&\Big [ \Lambda ^{\rm qf}_{A_1,E_1,t}(\Theta) + \Lambda^{\rm fis}_{A_1,E_1,t}(\Theta)\Big]P(Z_1,N_1,E_1,t).
\end{eqnarray}
The $W_{Z_1,N_1,Z'_1,N_1}$ ($W_{Z_1,N_1,Z_1,N'_1}$) was the mean transition probability from the channel ($Z_1,N_1,E_1$) to ($Z'_1,N_1,E'_1$) [or ($Z_1,N_1,E_1$) to ($Z_1,N'_1,E'_1$)]. The $d_{Z_1,N_1}$ are the microscopic dimension corresponding to the macroscopic state ($Z_1,N_1,E_1$). The sum contains all possible numbers of proton and neutron for the fragment $Z'_1$, $N'_1$ own. However, only one nucleon transfer at one time was supposed in the model with the relation $Z'_1$ = $Z_1$ $\pm$ 1, and $N'_1$ = $N_1$ $\pm$ 1. The excitation energy $E_1$ was the local excitation energy $\varepsilon^*_1$ in the fragment ($Z'_1$, $N'_1$)S \cite{PhysRevC.27.590}. The sticking time was evaluated by the deflection function \cite{LI1981107}. 
In the fusion process, the highly excited heavy fragments might lead to fission and the composite system may decay as quasifission. Both decay widths are calculated by the Kramers formula\cite{PhysRevC.68.034601,PhysRevC.27.2063}. 
Radial kinetic energy, angular momentum, and deformation relaxation time see Ref\cite{PhysRevC.80.057601}.
The local excitation energy is given by
\begin{eqnarray}
\varepsilon^* = E^{\rm diss} - (U_{\rm dr}(A_1,A_2) - U_{\rm dr}(A_{\rm P}, A_{\rm T}))
\end{eqnarray}
Where the $U_{\rm dr}(A_1, A_2)$ and $U_{\rm dr}(A_P, A_T)$ were the driving potentials of fragments $A_1$, $A_2$ and $A_{\rm P}$, $A_{\rm T}$, respectively. The excitation energy $E_{\rm x}$ of the composite system was converted from the relative kinetic energy dissipation\cite{PhysRevC.76.044606}. The potential energy surface (PES) of the DNS is written as
\begin{eqnarray}\label{pes}
&&U_{\rm dr}(A_1,A_2;J,\theta_1,\theta_2) =  B_1+B_2 - B_{CN}-V^{\rm CN}_{\rm rot}(J) \nonumber\\&&
+V_{\rm C}(A_1,A_2;\theta_1,\theta_2) + V_{\rm N}(A_1,A_2;\theta_1,\theta_2)
\end{eqnarray}

 Here $B_{i}$ ($i$ = 1, 2) and $B_{\rm CN}$ were the binding energies. The $V_{\rm rot}^{\rm CN}$ is the rotation energy of the compound nuclei. The $\beta_{\rm i}$ represent the quadrupole deformations of binary fragments. The $\theta_{\rm i}$ denotes collision orientations. The $V_{\rm C}$ and $V_{\rm N}$ are the Coulomb potential and nucleus-nucleus potential respectively.

The probability of all possible fragments is represented by solving a set of master equations. The hindrance in the fusion process is named inner fusion barrier $ B_{\rm fus}$ which is defined as the difference between the injection position and the B.G. point. These fragments overcome the inner barrier and may lead to fusion. Therefore, the fusion probability is evaluated by summing all of the fragments that could penetrate the inner fusion barrier. The fusion probability is evaluated by  

\begin{eqnarray}
P_{\rm CN}(E_{\rm c.m.},J,B)=\sum _{A=1}^{A _{\rm BG}} P(A,E_1,\tau_{\rm int}(E_{\rm c.m.},J,B)).
\end{eqnarray}

\subsection{ Survival probability}

The compound nuclei formed by all the nucleons transfer from projectile nuclei to target nuclei with certain excitation energies. The excited compound nuclei were extremely unstable which would be de-excited by evaporating $\gamma$-rays, neutrons, protons, $\alpha$ $etc.$) against fission. The survival probability of the channels x-th neutron, y-th proton and z-alpha. Development of a statistical evaporation model based on Weisskopf's evaporation theory.\cite{Chen_2016,FENG201082,FENG200933,Xin2021}
\begin{eqnarray}
&&W_{\rm sur}(E_{\rm CN}^*,x,y,z,J)=P(E_{\rm CN}^*,x,y,z,J) \times \nonumber\\&&  \prod_{i=1}^{x}\frac{\Gamma _n(E_i^*,J)}{\Gamma _{\rm tot}(E_i^*,J)} \prod_{j=1}^{y}\frac{\Gamma _p(E_j^*,J)}{\Gamma_{\rm tot}(E_i^*,J)} \prod_{k=1}^{z}\frac{\Gamma _\alpha (E_k^*,J)}{\Gamma _{\rm tot}(E_k^*,J)}, 
\end{eqnarray}
where the $E_{\rm CN}^*$ and $J$ were the excitation energy and the spin of the excited nucleus, respectively. The total width $\Gamma_{\rm tot}$ was the sum of partial widths of particle evaporation, $\gamma$-rays, and fission. The excitation energy $E_S^*$ before evaporating the $s$-th particles was evaluated by
\begin{eqnarray}
E_{s+1}^*=E_s^* - B _i ^n - B _j ^p - B_k ^\alpha - 2T_s
\end{eqnarray}
with the initial condition $E\rm_i^*$=$E_{\rm CN}^*$ and $s$=$i$+$j$+$k$. The $B_{\rm i}^n$, $B_{\rm j}^p$, $B_{\rm k}^\alpha$ are the separation energy of the $i$-th neutron, $j$-th proton, $k$-th alpha, respectively. The nuclear temperature $T_i$ was defined by $E_{\rm i}^*=\alpha T_{\rm i}^2-T_{\rm i}$ with the level density parameter $a$. The decay width of the $\gamma$-rays and the particle decay were evaluated with a similar method in Ref. \cite{Chen_2016}.
The $P(E_{\rm CN}^*,J)$ is the realization probability of evaporation channels.  

The $\Gamma _n(E_i^*,J)$, $\Gamma _p(E_i^*,J)$ and $\Gamma _{\alpha}(E_i^*,J)$ are the decay widths of particles n, p, $\alpha$, which are evaluated by the Weisskopf evaporation theory\cite{PhysRevC.68.014616}. The fission width $\Gamma_f(E^*,J)$ was calculated by the Bohr-Wheeler formula\cite{PhysRevC.80.057601,artza05,PhysRevC.76.044606}.

In our calculation, the fission barrier has a microscopic part and the macroscopic part which is written as
\begin{eqnarray}
B\rm_f(E^*,J)=B_f^{LD}+B_f^M(E^*=0,J)\rm exp(-E^*/E_D)
\end{eqnarray}
where the macroscopic part was derived from the liquid-drop model, as follows
\begin{eqnarray}
B^{\rm LD}\rm_f = \left \{\begin{array} {rl} 0.38(0.75 - x )E\rm_{s0} & , (1/3 < x < 2/3) \\ \\
0.83(1-x)^3 E\rm_{s0} & ,(2/3 < x < 1) \end{array} \right.
\end{eqnarray}
with 
\begin{eqnarray}
x=\frac{E\rm_{c0}}{2E\rm_{s0}}. 
\end{eqnarray}
Here, $E_{\rm c0}$ and $E_{\rm s0}$ were the surface energy and Coulomb energy of the spherical nuclear, respectively, which could be taken from the Myers-Swiatecki formula
\begin{eqnarray}
E_{\rm s0}=17.944[1-1.7826(\frac{N-Z}{A})^2]A^{2/3} \ \rm{MeV} 
\end{eqnarray}
and
\begin{eqnarray}
E_{\rm c0}=0.7053\frac{Z^2}{A^{1/3}} \  \rm{MeV}. 
\end{eqnarray}
Microcosmic shell correction energy was taken from \cite{Moller_1995}. Shell-damping energy was taken as $E_{\rm D}=5.48A^{1/3}/(1+1.3A^{-1/3})$ MeV.

\section{Results and discussion}\label{sec3}

In past decades, the dinuclear system model (DNS) can agree with the available experimental data well and own a strong predictive capability\cite{2011fennpr,PhysRevC.78.054607,FENG200650,nst2021niu,FENG2010384c,PhysRevC.90.014612,PhysRevC.100.011601,doi:10.1142/S021830130801091X,PhysRevC.76.044606,PhysRevC.80.057601,PhysRevC.89.037601,PhysRevC.85.041601,FZQ2009,Li_2006,Niu2021}. However, its predictive ability or the uncertainty of the prediction results depends strongly on the basic physical inputs, such as nuclear mass, quadrupole deformation, shell correction energy, fission barriers, barrier distribution and level density parameter and so on.
In this work, we focus on the effect of the nuclear mass model on the calculation of fusion probability and then discuss their effect on the cross section of superheavy nuclei.
Within the framework of the DNS, five type nuclear mass models (FRDM12, KTUY05, WS4, MS96, and HFB02) are used as inputs to systematically calculate the synthesis cross section of superheavy nuclei based on hot fusion reactions and to compare with the available experimental data.
We have analyzed in detail the three-stage capture-fusion-survival results for the $^{48}$Ca+$^{243}$Am$\rightarrow$$^{291}$Mc$^*$ reaction system. Finally, we have calculated in detail the synthesis of four projectile-target combinations of elements 119 and 120, namely $^{54}$Cr+$^{243}$Am, $^{55}$Mn+$^{243}$Am, $^{51}$V+$^{248}$Cm, and $^{54}$Cr+$^{248}$Cm. Among them, the IMP and the RIKEN chose $^{54}$Cr+$^{243}$Am and $^{51}$V+$^{248}$Cm respectively as combinations of projectile-target to synthesize element 119.
It is noteworthy that despite several years of experiments at RIKEN, no significant results have been achieved. However, there is hope on the horizon as the experiments at the IMP are scheduled to commence in the early part of this year. Encouragingly, preliminary test experiments involving hot-fusion reactions have already been conducted and have shown promising consistency with previous pioneering research\cite{arti22gan}.

\begin{figure}[htb]
\includegraphics[width=1.\linewidth]{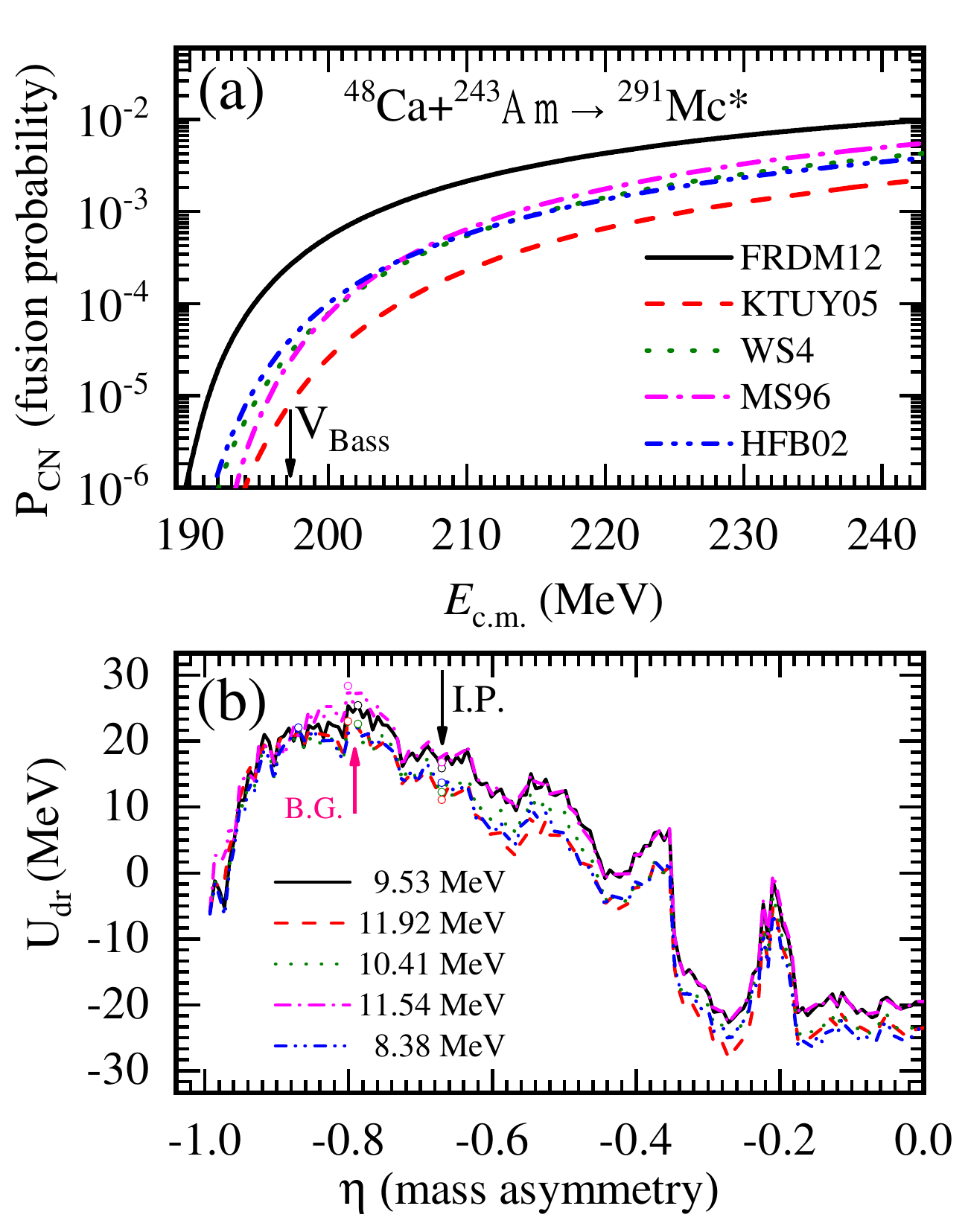}
\caption{\label{fig1} (Color online) In Figure (a), the fusion probability of the reaction system $^{48}$Ca+$^{243}$Am is calculated by solving the master equation based on five type nuclear mass models FRDM12, KTUY05, WS4, MS96, and HFB02, respectively. The five style lines and colors in each figure represent different nuclear mass models. In panel (b), the line with the arrow indicates the injection position of the projectile target and the bidirectional arrow indicates the value of the inner fusion barrier.}
\end{figure}

Mass is a pivotal characteristic of atomic nucleus, fundamental to both nuclear and astrophysical physics\cite{Kajino2023}. It encodes information about the shell structure, shape, and effective nucleon interactions, serving as a crucial input in numerous nuclear reactions\cite{Ming2022}. Furthermore, it holds the key to understanding the origin of cosmic elements. Since Weizsacker's proposal of the semi-empirical mass formula in 1935\cite{Weizsacker1935}, researchers have been relentlessly pursuing more precise nuclear mass models, encompassing diverse macro-microscopic frameworks, including the Strutinsky shell correction method\cite{STRUTINSKY1967420,STRUTINSKY19681}. As macro-microscopic nuclear mass models, the FRDM12, WS4, MS96, and KTUY05 exhibit varying extrapolation capabilities owing to their differing strengths of spin-orbit potential and deformation energies of nuclei.
The HFB02 represents a non-relativistic microscopic approach that relies on an effective nucleon-nucleon interaction of the Skyrme type.

Figure \ref{fig1} presents fusion probability and driver potential of the $^{48}$Ca+$^{243}$Am $\rightarrow$ $^{291}$Mc with the five type nuclear mass models.
Fusion probabilities of $^{48}$Ca+$^{243}$Am $\rightarrow$ $^{291}$Mc based on FRDM12, KTUY05, WS4, MS96, and HFB02 mass models are represented by a solid black line, a red dashed line, an olive green dotted line, a magenta dot-dashed line, and a blue double-dotted line in Fig. \ref{fig1}, respectively. 
From panel (a), the horizontal coordinate is the collision energy in the center-of-mass system and the vertical coordinate is the fusion probability. The arrow line stands for Bass potential.
The fusion probability increases exponentially with increasing collision energy, this is mainly because the large collision energies overcome the inner fusion barrier easily.
Below the Bass potential, fusion probabilities based on nuclear mass models of FRDM12, MS96, HFB02, WS4, and KTUY05 are in descending order. 
The fusion probabilities based on MS96, HFB02, and WS4 mass models are closed. 
The black arrow line and the red arrow line stand for the injection positions (I.P., $\eta$=0.67 for $^{48}$Ca+$^{243}$Am) and Businaro-Gallone region (B.G.) in panel (b), respectively.
The $\eta$ represents the mass asymmetry of the colliding system, which is defined as the ratio of the difference in masses $(A_{\rm P}-A_{\rm T})$ to the sum of masses $(A_{\rm P}+A_{\rm T})$, i.e., $\eta = (A_{\rm P}-A_{\rm T})/(A_{\rm P}+A_{\rm T})$. This formula quantifies the degree of asymmetry in the colliding system.
The inner fusion barrier is defined by the discrepancy value between the I.P. and the B.G. point, which serves as an indicator of the difficulty level associated with fusion.
The inner fusion barriers have been extracted from the potential energy surfaces (PESs) using the FRDM12, KTUY05, WS4, MS96, and HFB02 mass models, and their respective values are 9.53 MeV, 11.92 MeV, 10.41 MeV, 11.54 MeV, and 8.38 MeV.
The different mass models of nuclei could impact fusion probability directly, through the inner fusion barriers and B.G. positions.

\begin{figure}[htb]
\includegraphics[width=1.\linewidth]{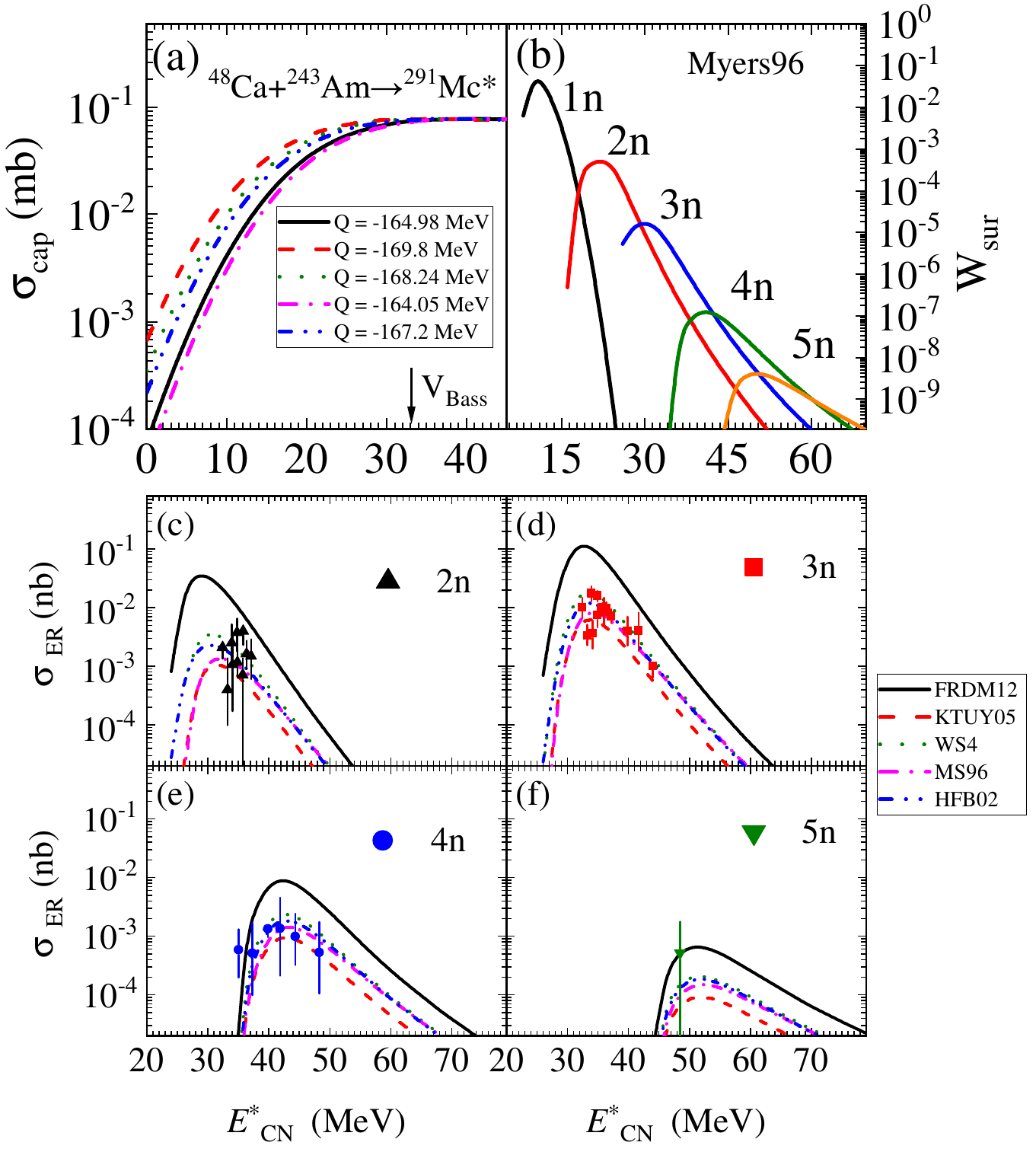}
\caption{\label{fig2}(Color online) Capture cross sections, survival probabilities of $^{48}$Ca+$^{243}$Am $\rightarrow$ $^{291}$Lv$^*$ and the evaporation residual cross sections compared with experimental data are presented. In panel (a), the horizontal coordinate is the excitation energy and the vertical coordinate is the capture cross sections. The results obtained based on different nuclear mass models are represented by different color lines. Panel (b) shows the survival probabilities calculated using the MS96 mass model. Panels (c)-(f) are the excitation functions of the evaporation residual cross sections of channels 2n-5n.}
\end{figure}

The influence of nuclear mass models on calculations of fusion probability has been discussed in Fig.\ref{fig1}.
To further investigate the effect of nuclear mass models on the residual evaporation cross-section, it is necessary to calculate the capture cross sections and survival probability, separately, and combine fusion probability. 
Figure \ref{fig2} presents the capture cross sections, survival probabilities, and evaporation residual cross sections of the $^{48}$Ca+$^{243}$Am $\rightarrow$ $^{291}$Mc$^*$, compared with experimental data.
For the reaction system $^{48}$Ca+$^{243}$Am $\rightarrow$ $^{291}$Mc$^*$, Q-values derived from the five type mass models of FRDM12, KTUY05, WS4, MS96, and HFB02 are -164.98 MeV, -169.8 MeV, -168.24 MeV, - 164.05 MeV, -167.2 MeV and their values differ within 4 MeV, as shown in panel (a), where the arrow line represent Bass potential. 
The calculation of the survival probability is using the MS96 mass model, as shown in panel (b), where different color lines stand for different neutron channels. These solid black, red, blue, olive, and yellow lines stand for 1n, 2n, 3n, 4n, and 5n, respectively.
 
Figure \ref{fig2} (c)-(f) shows excitation functions of the residual evaporation cross sections corresponding to 2n-5n channels, respectively. 
The experimental data are represented by black solid triangles, red squares, blue solid circles, and olive green solid inverted triangles corresponding to 2n, 3n, 4n, and 5n evaporation channels, respectively.
From panels (c) - (f), it was found that our calculations have good agreement with the experimental results, except the calculations using the FRDM12. The distribution region of excitation functions within the mass models of KTUY05, WS4, MS96, and HFB02 is less than one order of magnitude. 
And as can be seen from the analyses in Fig. \ref{fig2}(e) and (f), the calculated results of all mass show a better agreement with the experimental data,  The order of magnitude of the excitation function peaks is FRDM12, WS4, HFB02, MS96, and KTUY05 in the case of 4n and 5n channels.
It is noteworthy that the calculation of the capture cross section is independent of nuclear masses. In our calculations, the determination of survival probability solely relies on the MS96 mass model. Furthermore, five distinct nuclear models are employed in the computation of fusion probabilities. Evidently, nuclear mass models play a direct role in influencing the calculation of the synthesis cross sections for superheavy nuclei.

\begin{table}[htp]
\caption{Maximum values of evaporation residual cross sections (pb) for nine hot-fusion reactions calculated within five nuclear mass models and compared with experimental values.}
\label{table}
\begin{spacing}{1.3}
\setlength{\tabcolsep}{2.2pt}
\begin{tabular}{ccccccc}

\hline \hline
$^{48}$Ca+ & E(pb)     & F(pb)    & K(pb)    & W(pb)    & M(pb)    & H(pb)   \\ \hline
$^{238}$U (112)  & 2.46  & 17.3     & 0.93     & 1.88   & 2.21     & 5.13     \\
$^{237}$Np (113) & 0.87  & 16.4     & 0.73     & 2.4    & 1.43     & 1.2     \\
$^{242}$Pu (114) & 4.49  & 30.3     & 1.64     & 3.31   & 3.54     & 10.8    \\
$^{244}$Pu (114) & 10    & 26.1     & 1.49     & 3.44   & 5.61     & 5.82     \\
$^{243}$Am (115) & 17.48 & 19.5     & 11.5     & 21.7   & 8.75     & 26.3    \\
$^{245}$Cm (116) & 3.67  & 18.7     & 0.84     & 2.31   & 3.19     & 4.17   \\
$^{248}$Cm (116) & 3.41  & 2.47     & 0.53     & 0.19   & 2.97     & 2.38     \\
$^{249}$Bk (117) & 2.49  & 4.17     & 0.09     & 0.15   & 4.96     & 0.68     \\
$^{249}$Cf (118) & 0.56  & 0.13     & 0.04     & 0.09   & 0.52     & 0.03     \\ \hline
\hline

\end{tabular}
\end{spacing}
\end{table}

\begin{table}[htp]
\caption{The predicted cross sections for new SHEs with atomic numbers Z=119-120, in the neutron channels (2n-5n), were calculated for the reaction systems of $^{54}$Cr+$^{243}$Am, $^{55}$Mn+$^{243}$Am, $^{51}$V+$^{248}$Cm and $^{54}$Cr+$^{248}$Cm using five different nuclear mass models.}
\label{table2}
\begin{spacing}{1.3}
\setlength{\tabcolsep}{2.2pt}
\begin{tabular}{cccccc}

\hline \hline
 Reactions &  F(fb)     & K(fb)     & W(fb)    & M(fb)         & H(fb)   \\ \hline
$^{54}$Cr($^{243}$Am,2n)$^{295}$119 & 43       & 1.01     & 4.08   & 95       & 2.42     \\
$^{54}$Cr($^{243}$Am,3n)$^{294}$119 & 35       & 1.82     & 5.53   & 29.6     & 3.4     \\
$^{54}$Cr($^{243}$Am,4n)$^{293}$119 & 2.69     & 0.27     & 0.68   & 1.61     & 0.42    \\
$^{54}$Cr($^{243}$Am,5n)$^{292}$119 & 0.26     & 0.04     & 0.08   & 0.14     & 0.05    \\
\hline
$^{51}$V($^{248}$Cm,2n)$^{297}$119 & 2406      & 59.4     & 138.8   & 633.3     & 290.8     \\
$^{51}$V($^{248}$Cm,3n)$^{296}$119 & 1625      & 97.9     & 183.8   & 599.4     & 291.2    \\
$^{51}$V($^{248}$Cm,4n)$^{295}$119 & 58.19     & 7.74     & 12.3    & 28.94     & 15.4    \\
$^{51}$V($^{248}$Cm,5n)$^{294}$119 & 4.55      & 0.82     & 1.25    & 2.55      & 1.42    \\
\hline
$^{55}$Mn($^{243}$Am,2n)$^{296}$120 & 12.6     & 0.21     & 0.7     & 2.23     & 1.03    \\
$^{55}$Mn($^{243}$Am,3n)$^{295}$120 & 6.15     & 0.25     & 0.66    & 1.62     & 0.75    \\
$^{55}$Mn($^{243}$Am,4n)$^{294}$120 & 2.04     & 0.17     & 0.38    & 0.74     & 0.35    \\
$^{55}$Mn($^{243}$Am,5n)$^{293}$120 & 0.15     & 0.02     & 0.04    & 0.07     & 0.03    \\
\hline
$^{54}$Cr($^{248}$Cm,2n)$^{300}$120 & 14.5     & 0.19     & 0.62    & 1.45     & 0.63     \\
$^{54}$Cr($^{248}$Cm,3n)$^{300}$120 & 55.4     & 1.49     & 4.43    & 8.62     & 3.99     \\
$^{54}$Cr($^{248}$Cm,4n)$^{300}$120 & 16.16    & 1.09     & 2.73    & 4.34     & 2.2    \\
$^{54}$Cr($^{248}$Cm,5n)$^{300}$120 & 1.66     & 0.19     & 0.43    & 0.62     & 0.34   \\
\hline \hline
\end{tabular}
\end{spacing}
\end{table}
\begin{figure}[htb]
\includegraphics[width=1.\linewidth]{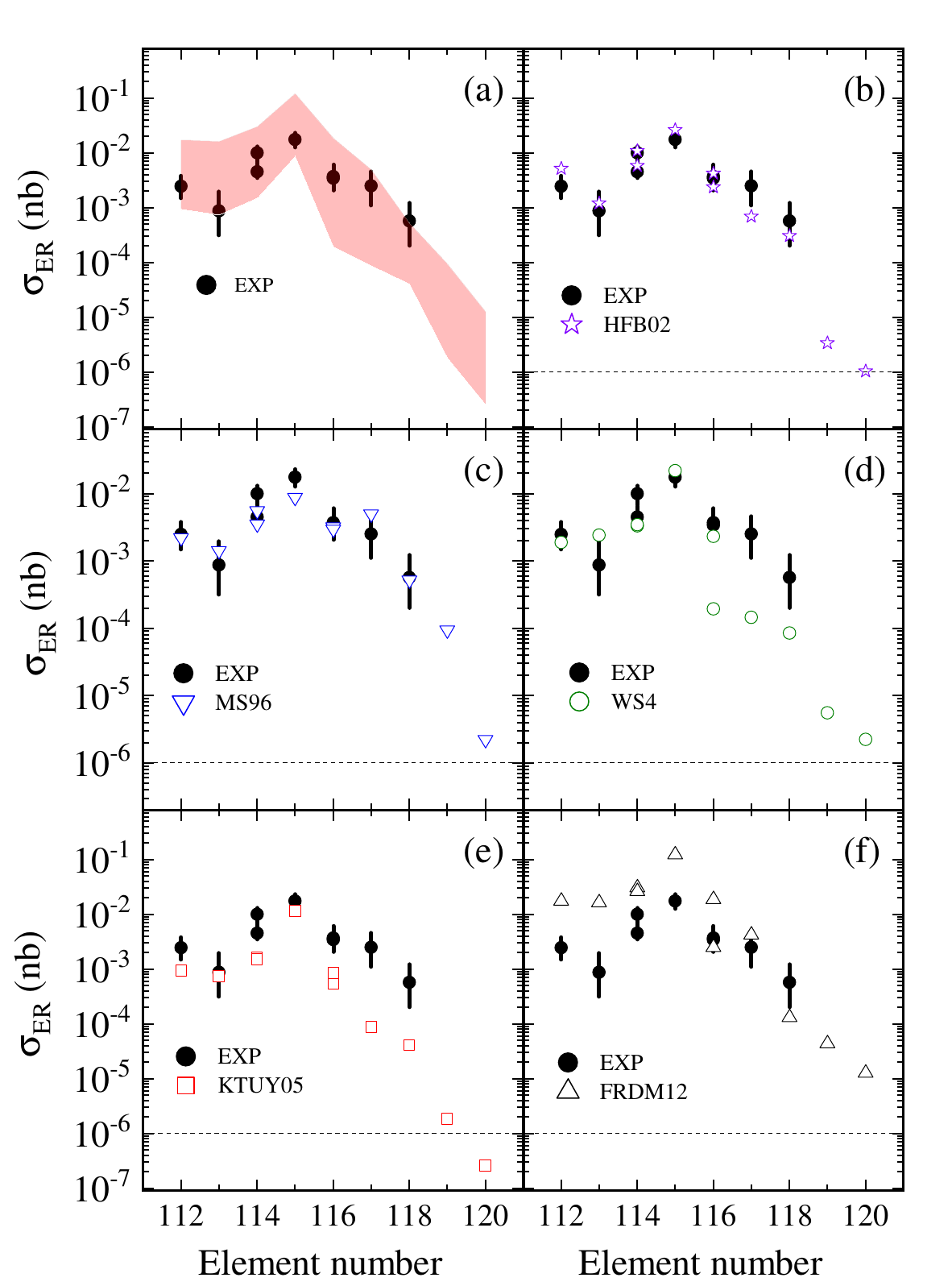}
\caption{\label{fig3}(Color online) 
Compare calculations to experimental data for the maximum value of evaporation residual cross sections of the hot-fusion reactions. 
The maximum value of synthesis cross sections of SHEs Z=119-120 based on the reaction systems of $^{54}$Cr+$^{243}$Am and $^{55}$Mn+$^{243}$Am are presented.
The pink shadow represents the distribution region of synthesis cross sections via the five types of nuclear mass models.
}
\end{figure}

To compare with all available hot-fusion experimental data, we have systematically calculated the reaction systems of $^{48}$Ca+$^{238}$U\cite{PhysRevC.70.064609}, $^{48}$Ca+$^{243}$Np\cite{doi:10.7566/JPSJ.86.085001,PhysRevC.72.034611}, $^{48}$Ca+$^{242,244}$Pu\cite{PhysRevLett.105.182701,PhysRevC.106.024612}, $^{48}$Ca+$^{243}$Am\cite{PhysRevC.69.021601}, $^{48}$Ca+$^{245,248}$Cm\cite{PhysRevC.69.054607}, $^{48}$Ca+$^{249}$Bk\cite{PhysRevLett.104.142502} and $^{48}$Ca+$^{249}$Cf\cite{PhysRevC.74.044602} based on five nuclear mass models. The maximum synthesis cross section for Z=112-118 is shown in Fig. \ref{fig3} and the detailed data is listed in table \ref{table}.

Figure \ref{fig3} shows the comparison of experimental results and calculations.
The experimental data are marked by black solid circles with error bars.
The pink shadow indicates the distribution region of the largest predictions of synthesis cross section of SHEs Z=112-120 via the five types of nuclear mass models, as shown in panel (a). 
It was found that the experimental data falls within the pink shadow, indicating that our calculations are consistent with the experimental results.
The experimental data distribution of the synthesis SHEs with atomic numbers ranging from 112 to 118 exhibits a trend of first increasing and then decreasing. The peak value occurs at Z=115, with a magnitude of 17.48 pb.
The predicted distribution region for the cross section of the SHE Z=119 via the reaction system $^{54}$Cr+$^{243}$Am $\rightarrow$ $^{297}$119 is between 1.8 and 95 fb, and the variations among different mass models are less than one order of magnitude.
The reaction system $^{55}$Mn+$^{243}$Am$\rightarrow$ $^{298}$120 is utilized for the synthesis of the SHE with atomic number Z=120, and the predicted range for the synthesis cross section is between 0.25 and 12 fb.
The black hollow triangles, red hollow squares, olive hollow circles, blue hollow inverted triangles, and purple hollow pentagrams represent the maximum values of evaporation residual cross sections obtained using the FRDM12, KTUY05, WS4, MS96, and HFB02 mass models, respectively, as listed in panels (b) to (f).
It was found that the calculations employing the mass models of HFB02 and MS96 exhibit better agreement with experimental data in panels (b) and (e) compared to those obtained using other nuclear mass models. On the other hand, the calculations utilizing the mass models of WS4 and KTUY05 slightly underestimate the experimental results in panels (d) and (e). Furthermore, the predictions based on the mass model of FRDM12 slightly overestimate the experimental data in panel (f).
In table \ref{table}, the letters F, K, W, M, and H represent the nuclear mass models FRDM12, KTUY05, WS4, MS96, and HFB02, respectively. The letter E is corresponding to experimental data. 


\begin{figure}
  \centering
  \includegraphics[width=0.5\textwidth]{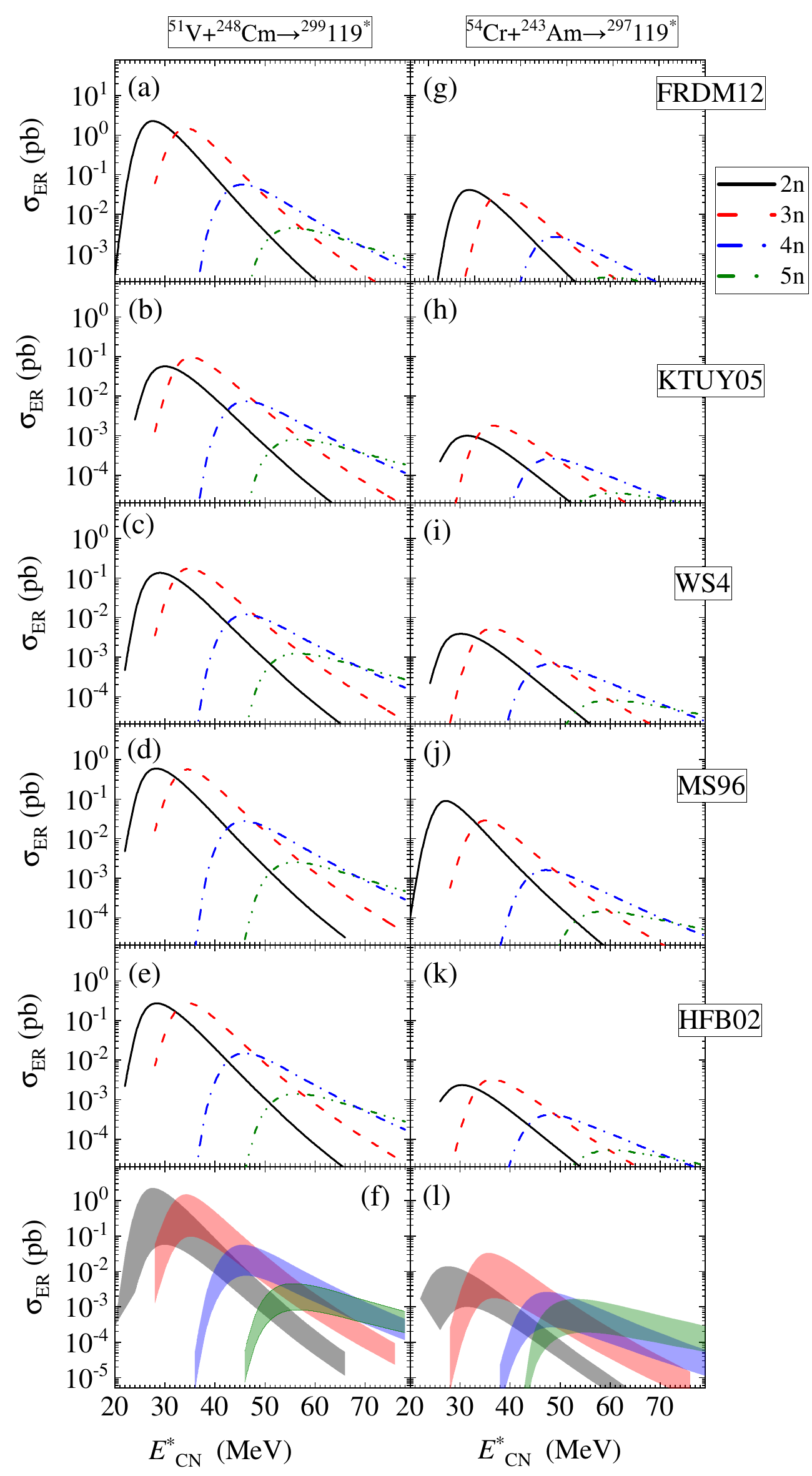}
  \caption{\label{fig4}(Color online) The excitation functions of residual evaporation cross section for the synthesis of element 119 via the projectile-target combinations of $^{51}$V+$^{248}$Cm, $^{54}$Cr+$^{243}$Am, respectively. The calculations using five types of nuclear mass models are listed in panels (a)-(e) and (g)-(k). The different neutron evaporation channels are indicated by different color line patterns. The color shadows stand for the distribution region of excitation functions.}
\end{figure}

To predict the synthesis cross sections of new SHEs with atomic numbers Z=119 and evaluate the influence of different nuclear mass models on the predicted results, the calculations of the reaction systems of $^{51}$V+$^{248}$Am $\rightarrow$ $^{299}$119$^*$ and $^{54}$Cr+$^{243}$Am $\rightarrow$ $^{297}$119$^*$ have been performed using the five nuclear mass models systematically.
Figure \ref{fig4} present the excitation functions of evaporation residual cross sections of $^{51}$V+$^{248}$Am $\rightarrow$ $^{299}$119$^*$, $^{54}$Cr+$^{243}$Am $\rightarrow$ $^{297}$119$^*$, which correspond to the left column of panels and the right column of panels respectively. 
The calculations based on the nuclear mass models of FRDM12, KTUY05, WS4, MS96, and HFB02 are represented by the panels from the first row to the fifth row, respectively.
The black solid lines, red dash lines, blue dot-dash lines, and olive green dote-dote dash lines represent the 2n, 3n, 4n, and 5n channels, respectively.
The one-neutron channels are too small to be ignored in panels. 
It was found that the results obtained from the $^{51}$V+$^{248}$Am $\rightarrow$ $^{299}$119$^*$ are larger that that obtained from the $^{54}$Cr+$^{243}$Am $\rightarrow$ $^{297}$119$^*$, because the former has larger fusion probability.
The accuracy peak value of each channel for figure \ref{fig4} has been listed in table \ref{table2}.
The shadows of black, red, blue, and olive denote the 2n, 3n, 4n, and 5n channels of the calculated distribution regions of excitation functions for evaporation residual cross sections using the five nuclear mass models in the panels (f) and (l). 
The utilization of different nuclear mass models significantly impacts the accuracy and reliability of the calculations.
The calculation error regions resulting from different nuclear mass models are less than one order of magnitude.
The mass model FRDM12 produces the largest cross sections, whereas the smallest results are obtained from the mass model KTUY05.
In our calculations, the optimal evaporation channels, excitation energies, and synthesis cross sections for the $^{51}$V+$^{248}$Am $\rightarrow$ $^{299}$119$^*$ are two-neutron channel, 28-30 MeV, and 0.06-2.4 pb, respectively. And that for the $^{54}$Cr+$^{243}$Am $\rightarrow$ $^{297}$119$^*$ are three-neutron channel, 36 MeV, and 1.8-34.9 fb.


\begin{figure}
  \centering
  \includegraphics[width=0.5\textwidth]{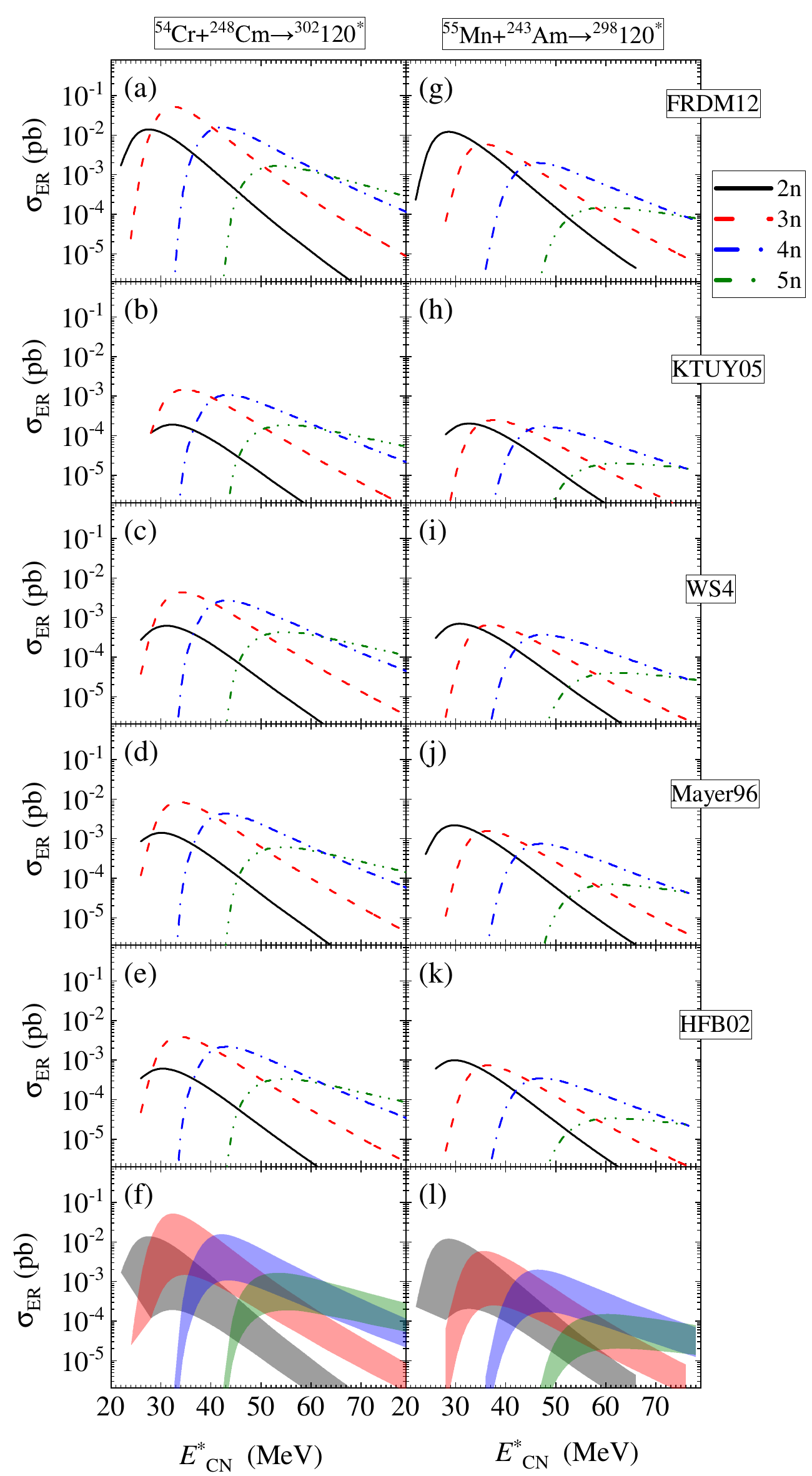}
  \caption{ The excitation functions of residual evaporation cross section for the synthesis of element 120 via the projectile-target combinations of $^{54}$Cr+$^{248}$Cm, $^{55}$Mn+$^{243}$Am, respectively. The calculations using five types of nuclear mass models are listed in panels (a)-(e) and (g)-(k). The different neutron evaporation channels are indicated by different color line patterns. The color shadows stand for the distribution region of excitation functions.}
  \label{fig5}
\end{figure}

In order to predict the synthesis cross sections of new superheavy elements (SHEs) with atomic numbers Z=120 and assess the impact of various nuclear mass models on the anticipated outcomes, systematic calculations were conducted utilizing five distinct nuclear mass models for the reaction systems of $^{54}$Cr+$^{248}$Cm $\rightarrow$ $^{302}$120$^*$ and $^{55}$Ca+$^{243}$Am $\rightarrow$ $^{298}$120$^*$, as shown in Fig.\ref{fig5}.
Figure \ref{fig5} is similar to Fig. \ref{fig4}, but for $^{54}$Cr+$^{248}$Cm $\rightarrow$ $^{302}$120$^*$ and $^{55}$Ca+$^{243}$Am $\rightarrow$ $^{298}$120$^*$.
From Fig. \ref{fig5}, the utilization of various nuclear mass models has a substantial influence on the precision and dependability of the calculations. Notably, the calculation error margins stemming from diverse nuclear mass models are less than one order of magnitude. Among the models considered, the FRDM12 mass model consistently yields the largest cross sections, while the smallest outcomes are consistently generated by the KTUY05 mass model.
The peak accuracy values corresponding to each channel depicted in Figure \ref{fig5} have been comprehensively outlined in Table \ref{table2}.
The optimal evaporation channel, excitation energy range, and synthesis cross section for the reaction $^{54}$Cr+$^{248}$Cm $\rightarrow$ $^{302}$120$^*$ are the three-neutron channel, 34-36 MeV, and 1.4-55.4 fb, respectively. 
Similarly, for the reaction $^{55}$Mn+$^{243}$Am $\rightarrow$ $^{298}$120$^*$, the optimal evaporation channel is the two-neutron channel, with an excitation energy of 30-32 MeV, and a synthesis cross sections ranging from 0.2 to 11.6 fb.

\section{Conclusions}\label{sec4}

The atomic nuclear mass serves as a crucial fundamental physical parameter in the DNS model.
Different nuclear mass models exhibit significant variations not only in their predictions of known nuclear masses but also in their extrapolation capabilities, leading to substantial discrepancies in Q-value and particle separation energies.
The calculations utilizing five nuclear mass models specifically, FRDM12, KTUY05, WS4, MS96, and HFB02, for hot-fusion reactions, have undergone comprehensive analysis. 
A rigorous comparative evaluation has been conducted between these computed results and available experimental data to ensure their accuracy and reliability.
Different nuclear mass models are crucial in the calculations of fusion probability and survival probability. 
In this paper, our primary focus is to delve into the influence of various nuclear mass models on the computation of fusion probability within the DNS model. We aim to provide a comprehensive analysis of how these models affect fusion probability calculations and their subsequent implications for nuclear reaction studies. 
In this paper, the statistical model based on the MS96 mass model is employed for all calculations to determine the survival probability of the superheavy compound nuclei.

Firstly, in the calculations of the fusion-evaporation reaction $^{48}$Ca+$^{243}$Am $\rightarrow$ $^{291}$Mc, it was observed that distinct nuclear mass models yield varying inner fusion barriers and differing positions of the B.G. point. These disparities result in diverse fusion probabilities, emphasizing the significant influence of the chosen nuclear mass model on fusion reaction outcomes. 
The synthesis cross-sections obtained using different mass models are distributed within an order of magnitude.
The model error of the DNS model for calculating the synthesis cross-section of superheavy nuclei is given by utilizing the uncertainty of nuclear mass.
In comparison with the available hot-fusion experimental data, our calculations exhibit a good agreement, with the error range falling within an order of magnitude.
Prediction of synthesis cross sections for new SHEs with Z=119-120 involved systematic calculations for four reactions. For $^{51}$V+$^{248}$Am, the optimal outcomes are the two-neutron channel, 28-30 MeV excitation, and 0.06-2.4 pb cross section. For $^{54}$Cr+$^{243}$Am, they are the three-neutron channel, 36 MeV excitation, and 1.8-34.9 fb cross section. Similarly, for $^{54}$Cr+$^{248}$Cm, the best results are the three-neutron channel, 34-36 MeV, and 1.4-55.4 fb, while for $^{55}$Mn+$^{243}$Am, they are the two-neutron channel, 30-32 MeV, and 0.2-11.6 fb.

\section{Acknowledgements}

This work was supported by National Science Foundation of China (NSFC) (Grants No. 12105241,12175072), NSF of Jiangsu Province (Grants No. BK20210788) and University Science Research Project of Jiangsu Province (Grants No. 21KJB140026). 


\begin{thebibliography}{80}%
\makeatletter
\providecommand \@ifxundefined [1]{%
 \@ifx{#1\undefined}
}%
\providecommand \@ifnum [1]{%
 \ifnum #1\expandafter \@firstoftwo
 \else \expandafter \@secondoftwo
 \fi
}%
\providecommand \@ifx [1]{%
 \ifx #1\expandafter \@firstoftwo
 \else \expandafter \@secondoftwo
 \fi
}%
\providecommand \natexlab [1]{#1}%
\providecommand \enquote  [1]{``#1''}%
\providecommand \bibnamefont  [1]{#1}%
\providecommand \bibfnamefont [1]{#1}%
\providecommand \citenamefont [1]{#1}%
\providecommand \href@noop [0]{\@secondoftwo}%
\providecommand \href [0]{\begingroup \@sanitize@url \@href}%
\providecommand \@href[1]{\@@startlink{#1}\@@href}%
\providecommand \@@href[1]{\endgroup#1\@@endlink}%
\providecommand \@sanitize@url [0]{\catcode `\\12\catcode `\$12\catcode
  `\&12\catcode `\#12\catcode `\^12\catcode `\_12\catcode `\%12\relax}%
\providecommand \@@startlink[1]{}%
\providecommand \@@endlink[0]{}%
\providecommand \url  [0]{\begingroup\@sanitize@url \@url }%
\providecommand \@url [1]{\endgroup\@href {#1}{\urlprefix }}%
\providecommand \urlprefix  [0]{URL }%
\providecommand \Eprint [0]{\href }%
\providecommand \doibase [0]{http://dx.doi.org/}%
\providecommand \selectlanguage [0]{\@gobble}%
\providecommand \bibinfo  [0]{\@secondoftwo}%
\providecommand \bibfield  [0]{\@secondoftwo}%
\providecommand \translation [1]{[#1]}%
\providecommand \BibitemOpen [0]{}%
\providecommand \bibitemStop [0]{}%
\providecommand \bibitemNoStop [0]{.\EOS\space}%
\providecommand \EOS [0]{\spacefactor3000\relax}%
\providecommand \BibitemShut  [1]{\csname bibitem#1\endcsname}%
\let\auto@bib@innerbib\@empty
\bibitem [{\citenamefont {Thoennessen}(2013)}]{THOENNESSEN2013312}%
  \BibitemOpen
  \bibfield  {author} {\bibinfo {author} {\bibfnamefont {M.}~\bibnamefont
  {Thoennessen}},\ }\href {\doibase https://doi.org/10.1016/j.adt.2012.03.003}
  {\bibfield  {journal} {\bibinfo  {journal} {Atom. Data Nucl. Data Tab.}\
  }\textbf {\bibinfo {volume} {99}},\ \bibinfo {pages} {312} (\bibinfo {year}
  {2013})}\BibitemShut {NoStop}%
\bibitem [{\citenamefont {Tarasov}\ \emph {et~al.}(2024)\citenamefont
  {Tarasov}, \citenamefont {Gade}, \citenamefont {Fukushima}, \citenamefont
  {Hausmann}, \citenamefont {Kwan}, \citenamefont {Portillo}, \citenamefont
  {Smith}, \citenamefont {Ahn}, \citenamefont {Bazin}, \citenamefont {Chyzh},
  \citenamefont {Giraud}, \citenamefont {Haak}, \citenamefont {Kubo},
  \citenamefont {Morrissey}, \citenamefont {Ostroumov}, \citenamefont
  {Richardson}, \citenamefont {Sherrill}, \citenamefont {Stolz}, \citenamefont
  {Watters}, \citenamefont {Weisshaar},\ and\ \citenamefont
  {Zhang}}]{PhysRevLett.132.072501}%
  \BibitemOpen
  \bibfield  {author} {\bibinfo {author} {\bibfnamefont {O.~B.}\ \bibnamefont
  {Tarasov}}, \bibinfo {author} {\bibfnamefont {A.}~\bibnamefont {Gade}},
  \bibinfo {author} {\bibfnamefont {K.}~\bibnamefont {Fukushima}}, \bibinfo
  {author} {\bibfnamefont {M.}~\bibnamefont {Hausmann}}, \bibinfo {author}
  {\bibfnamefont {E.}~\bibnamefont {Kwan}}, \bibinfo {author} {\bibfnamefont
  {M.}~\bibnamefont {Portillo}}, \bibinfo {author} {\bibfnamefont
  {M.}~\bibnamefont {Smith}}, \bibinfo {author} {\bibfnamefont {D.~S.}\
  \bibnamefont {Ahn}}, \bibinfo {author} {\bibfnamefont {D.}~\bibnamefont
  {Bazin}}, \bibinfo {author} {\bibfnamefont {R.}~\bibnamefont {Chyzh}},
  \bibinfo {author} {\bibfnamefont {S.}~\bibnamefont {Giraud}}, \bibinfo
  {author} {\bibfnamefont {K.}~\bibnamefont {Haak}}, \bibinfo {author}
  {\bibfnamefont {T.}~\bibnamefont {Kubo}}, \bibinfo {author} {\bibfnamefont
  {D.~J.}\ \bibnamefont {Morrissey}}, \bibinfo {author} {\bibfnamefont {P.~N.}\
  \bibnamefont {Ostroumov}}, \bibinfo {author} {\bibfnamefont {I.}~\bibnamefont
  {Richardson}}, \bibinfo {author} {\bibfnamefont {B.~M.}\ \bibnamefont
  {Sherrill}}, \bibinfo {author} {\bibfnamefont {A.}~\bibnamefont {Stolz}},
  \bibinfo {author} {\bibfnamefont {S.}~\bibnamefont {Watters}}, \bibinfo
  {author} {\bibfnamefont {D.}~\bibnamefont {Weisshaar}}, \ and\ \bibinfo
  {author} {\bibfnamefont {T.}~\bibnamefont {Zhang}},\ }\href {\doibase
  10.1103/PhysRevLett.132.072501} {\bibfield  {journal} {\bibinfo  {journal}
  {Phys. Rev. Lett.}\ }\textbf {\bibinfo {volume} {132}},\ \bibinfo {pages}
  {072501} (\bibinfo {year} {2024})}\BibitemShut {NoStop}%
\bibitem [{\citenamefont {Yang}\ \emph {et~al.}(2024)\citenamefont {Yang},
  \citenamefont {Gan}, \citenamefont {Li}, \citenamefont {Liu}, \citenamefont
  {Xu}, \citenamefont {Liu}, \citenamefont {Zhang}, \citenamefont {Zhang},
  \citenamefont {Huang}, \citenamefont {Yuan}, \citenamefont {Wang},
  \citenamefont {Ma}, \citenamefont {Wang}, \citenamefont {Han}, \citenamefont
  {Rohilla}, \citenamefont {Zuo}, \citenamefont {Xiao}, \citenamefont {Zhang},
  \citenamefont {Zhu}, \citenamefont {Yue}, \citenamefont {Tian}, \citenamefont
  {Wang}, \citenamefont {Yang}, \citenamefont {Zhao}, \citenamefont {Huang},
  \citenamefont {Li}, \citenamefont {Sun}, \citenamefont {Wang}, \citenamefont
  {Yang}, \citenamefont {Lu}, \citenamefont {Yang}, \citenamefont {Zhou},
  \citenamefont {Huang}, \citenamefont {Wang}, \citenamefont {Zhou},
  \citenamefont {Ren},\ and\ \citenamefont {Xu}}]{PhysRevLett.132.072502}%
  \BibitemOpen
  \bibfield  {author} {\bibinfo {author} {\bibfnamefont {H.~B.}\ \bibnamefont
  {Yang}}, \bibinfo {author} {\bibfnamefont {Z.~G.}\ \bibnamefont {Gan}},
  \bibinfo {author} {\bibfnamefont {Y.~J.}\ \bibnamefont {Li}}, \bibinfo
  {author} {\bibfnamefont {M.~L.}\ \bibnamefont {Liu}}, \bibinfo {author}
  {\bibfnamefont {S.~Y.}\ \bibnamefont {Xu}}, \bibinfo {author} {\bibfnamefont
  {C.}~\bibnamefont {Liu}}, \bibinfo {author} {\bibfnamefont {M.~M.}\
  \bibnamefont {Zhang}}, \bibinfo {author} {\bibfnamefont {Z.~Y.}\ \bibnamefont
  {Zhang}}, \bibinfo {author} {\bibfnamefont {M.~H.}\ \bibnamefont {Huang}},
  \bibinfo {author} {\bibfnamefont {C.~X.}\ \bibnamefont {Yuan}}, \bibinfo
  {author} {\bibfnamefont {S.~Y.}\ \bibnamefont {Wang}}, \bibinfo {author}
  {\bibfnamefont {L.}~\bibnamefont {Ma}}, \bibinfo {author} {\bibfnamefont
  {J.~G.}\ \bibnamefont {Wang}}, \bibinfo {author} {\bibfnamefont {X.~C.}\
  \bibnamefont {Han}}, \bibinfo {author} {\bibfnamefont {A.}~\bibnamefont
  {Rohilla}}, \bibinfo {author} {\bibfnamefont {S.~Q.}\ \bibnamefont {Zuo}},
  \bibinfo {author} {\bibfnamefont {X.}~\bibnamefont {Xiao}}, \bibinfo {author}
  {\bibfnamefont {X.~B.}\ \bibnamefont {Zhang}}, \bibinfo {author}
  {\bibfnamefont {L.}~\bibnamefont {Zhu}}, \bibinfo {author} {\bibfnamefont
  {Z.~F.}\ \bibnamefont {Yue}}, \bibinfo {author} {\bibfnamefont {Y.~L.}\
  \bibnamefont {Tian}}, \bibinfo {author} {\bibfnamefont {Y.~S.}\ \bibnamefont
  {Wang}}, \bibinfo {author} {\bibfnamefont {C.~L.}\ \bibnamefont {Yang}},
  \bibinfo {author} {\bibfnamefont {Z.}~\bibnamefont {Zhao}}, \bibinfo {author}
  {\bibfnamefont {X.~Y.}\ \bibnamefont {Huang}}, \bibinfo {author}
  {\bibfnamefont {Z.~C.}\ \bibnamefont {Li}}, \bibinfo {author} {\bibfnamefont
  {L.~C.}\ \bibnamefont {Sun}}, \bibinfo {author} {\bibfnamefont {J.~Y.}\
  \bibnamefont {Wang}}, \bibinfo {author} {\bibfnamefont {H.~R.}\ \bibnamefont
  {Yang}}, \bibinfo {author} {\bibfnamefont {Z.~W.}\ \bibnamefont {Lu}},
  \bibinfo {author} {\bibfnamefont {W.~Q.}\ \bibnamefont {Yang}}, \bibinfo
  {author} {\bibfnamefont {X.~H.}\ \bibnamefont {Zhou}}, \bibinfo {author}
  {\bibfnamefont {W.~X.}\ \bibnamefont {Huang}}, \bibinfo {author}
  {\bibfnamefont {N.}~\bibnamefont {Wang}}, \bibinfo {author} {\bibfnamefont
  {S.~G.}\ \bibnamefont {Zhou}}, \bibinfo {author} {\bibfnamefont {Z.~Z.}\
  \bibnamefont {Ren}}, \ and\ \bibinfo {author} {\bibfnamefont {H.~S.}\
  \bibnamefont {Xu}},\ }\href {\doibase 10.1103/PhysRevLett.132.072502}
  {\bibfield  {journal} {\bibinfo  {journal} {Phys. Rev. Lett.}\ }\textbf
  {\bibinfo {volume} {132}},\ \bibinfo {pages} {072502} (\bibinfo {year}
  {2024})}\BibitemShut {NoStop}%
\bibitem [{\citenamefont {Heinlein}\ \emph {et~al.}(1978)\citenamefont
  {Heinlein}, \citenamefont {Bachmann},\ and\ \citenamefont
  {Rudolph}}]{HEINLEIN1978407}%
  \BibitemOpen
  \bibfield  {author} {\bibinfo {author} {\bibfnamefont {G.}~\bibnamefont
  {Heinlein}}, \bibinfo {author} {\bibfnamefont {K.}~\bibnamefont {Bachmann}},
  \ and\ \bibinfo {author} {\bibfnamefont {J.}~\bibnamefont {Rudolph}},\ }in\
  \href {\doibase https://doi.org/10.1016/B978-0-08-022946-1.50051-7} {\emph
  {\bibinfo {booktitle} {Superheavy Elements}}},\ \bibinfo {editor} {edited by\
  \bibinfo {editor} {\bibfnamefont {M.}~\bibnamefont {Lodhi}}}\ (\bibinfo
  {publisher} {Pergamon},\ \bibinfo {year} {1978})\ pp.\ \bibinfo {pages}
  {407--414}\BibitemShut {NoStop}%
\bibitem [{\citenamefont {Flerov}\ \emph {et~al.}(1970)\citenamefont {Flerov},
  \citenamefont {Oganesyan}, \citenamefont {Lobanov}, \citenamefont {Lazarev},
  \citenamefont {Tretyakova}, \citenamefont {Kolesov},\ and\ \citenamefont
  {Plotko}}]{1970flog}%
  \BibitemOpen
  \bibfield  {author} {\bibinfo {author} {\bibfnamefont {G.~N.}\ \bibnamefont
  {Flerov}}, \bibinfo {author} {\bibfnamefont {Y.~T.}\ \bibnamefont
  {Oganesyan}}, \bibinfo {author} {\bibfnamefont {Y.~V.}\ \bibnamefont
  {Lobanov}}, \bibinfo {author} {\bibfnamefont {Y.~A.}\ \bibnamefont
  {Lazarev}}, \bibinfo {author} {\bibfnamefont {S.~P.}\ \bibnamefont
  {Tretyakova}}, \bibinfo {author} {\bibfnamefont {I.~V.}\ \bibnamefont
  {Kolesov}}, \ and\ \bibinfo {author} {\bibfnamefont {V.~M.}\ \bibnamefont
  {Plotko}},\ }\href@noop {} {\bibfield  {journal} {\bibinfo  {journal} {Sov.
  At. Energy}\ }\textbf {\bibinfo {volume} {29}},\ \bibinfo {pages} {967}
  (\bibinfo {year} {1970})}\BibitemShut {NoStop}%
\bibitem [{\citenamefont {G.}\ \emph {et~al.}(1970)\citenamefont {G.},
  \citenamefont {N.}, \citenamefont {E.}, \citenamefont {H.},\ and\
  \citenamefont {E.}}]{PhysRevLett.24.1498}%
  \BibitemOpen
  \bibfield  {author} {\bibinfo {author} {\bibfnamefont {A.}~\bibnamefont
  {G.}}, \bibinfo {author} {\bibfnamefont {M.}~\bibnamefont {N.}}, \bibinfo
  {author} {\bibfnamefont {K.}~\bibnamefont {E.}}, \bibinfo {author}
  {\bibfnamefont {J.}~\bibnamefont {H.}}, \ and\ \bibinfo {author}
  {\bibfnamefont {P.}~\bibnamefont {E.}},\ }\href {\doibase
  10.1103/PhysRevLett.24.1498} {\bibfield  {journal} {\bibinfo  {journal}
  {Phys. Rev. Lett.}\ }\textbf {\bibinfo {volume} {24}},\ \bibinfo {pages}
  {1498} (\bibinfo {year} {1970})}\BibitemShut {NoStop}%
\bibitem [{\citenamefont {Ghiorso}\ \emph {et~al.}(1974)\citenamefont
  {Ghiorso}, \citenamefont {Nitschke}, \citenamefont {Alonso}, \citenamefont
  {Alonso}, \citenamefont {Nurmia}, \citenamefont {Seaborg}, \citenamefont
  {Hulet},\ and\ \citenamefont {Lougheed}}]{PhysRevLett.33.1490}%
  \BibitemOpen
  \bibfield  {author} {\bibinfo {author} {\bibfnamefont {A.}~\bibnamefont
  {Ghiorso}}, \bibinfo {author} {\bibfnamefont {J.~M.}\ \bibnamefont
  {Nitschke}}, \bibinfo {author} {\bibfnamefont {J.~R.}\ \bibnamefont
  {Alonso}}, \bibinfo {author} {\bibfnamefont {C.~T.}\ \bibnamefont {Alonso}},
  \bibinfo {author} {\bibfnamefont {M.}~\bibnamefont {Nurmia}}, \bibinfo
  {author} {\bibfnamefont {G.~T.}\ \bibnamefont {Seaborg}}, \bibinfo {author}
  {\bibfnamefont {E.~K.}\ \bibnamefont {Hulet}}, \ and\ \bibinfo {author}
  {\bibfnamefont {R.~W.}\ \bibnamefont {Lougheed}},\ }\href {\doibase
  10.1103/PhysRevLett.33.1490} {\bibfield  {journal} {\bibinfo  {journal}
  {Phys. Rev. Lett.}\ }\textbf {\bibinfo {volume} {33}},\ \bibinfo {pages}
  {1490} (\bibinfo {year} {1974})}\BibitemShut {NoStop}%
\bibitem [{\citenamefont {Ellison}\ \emph {et~al.}(2010)\citenamefont
  {Ellison}, \citenamefont {Gregorich}, \citenamefont {Berryman}, \citenamefont
  {Bleuel}, \citenamefont {Clark}, \citenamefont
  {Dragojevi\ifmmode~\acute{c}\else \'{c}\fi{}}, \citenamefont {Dvorak},
  \citenamefont {Fallon}, \citenamefont {Fineman-Sotomayor}, \citenamefont
  {Gates}, \citenamefont {Gothe}, \citenamefont {Lee}, \citenamefont
  {Loveland}, \citenamefont {McLaughlin}, \citenamefont {Paschalis},
  \citenamefont {Petri}, \citenamefont {Qian}, \citenamefont {Stavsetra},
  \citenamefont {Wiedeking},\ and\ \citenamefont
  {Nitsche}}]{PhysRevLett.105.182701}%
  \BibitemOpen
  \bibfield  {author} {\bibinfo {author} {\bibfnamefont {P.~A.}\ \bibnamefont
  {Ellison}}, \bibinfo {author} {\bibfnamefont {K.~E.}\ \bibnamefont
  {Gregorich}}, \bibinfo {author} {\bibfnamefont {J.~S.}\ \bibnamefont
  {Berryman}}, \bibinfo {author} {\bibfnamefont {D.~L.}\ \bibnamefont
  {Bleuel}}, \bibinfo {author} {\bibfnamefont {R.~M.}\ \bibnamefont {Clark}},
  \bibinfo {author} {\bibfnamefont {I.}~\bibnamefont
  {Dragojevi\ifmmode~\acute{c}\else \'{c}\fi{}}}, \bibinfo {author}
  {\bibfnamefont {J.}~\bibnamefont {Dvorak}}, \bibinfo {author} {\bibfnamefont
  {P.}~\bibnamefont {Fallon}}, \bibinfo {author} {\bibfnamefont
  {C.}~\bibnamefont {Fineman-Sotomayor}}, \bibinfo {author} {\bibfnamefont
  {J.~M.}\ \bibnamefont {Gates}}, \bibinfo {author} {\bibfnamefont {O.~R.}\
  \bibnamefont {Gothe}}, \bibinfo {author} {\bibfnamefont {I.~Y.}\ \bibnamefont
  {Lee}}, \bibinfo {author} {\bibfnamefont {W.~D.}\ \bibnamefont {Loveland}},
  \bibinfo {author} {\bibfnamefont {J.~P.}\ \bibnamefont {McLaughlin}},
  \bibinfo {author} {\bibfnamefont {S.}~\bibnamefont {Paschalis}}, \bibinfo
  {author} {\bibfnamefont {M.}~\bibnamefont {Petri}}, \bibinfo {author}
  {\bibfnamefont {J.}~\bibnamefont {Qian}}, \bibinfo {author} {\bibfnamefont
  {L.}~\bibnamefont {Stavsetra}}, \bibinfo {author} {\bibfnamefont
  {M.}~\bibnamefont {Wiedeking}}, \ and\ \bibinfo {author} {\bibfnamefont
  {H.}~\bibnamefont {Nitsche}},\ }\href {\doibase
  10.1103/PhysRevLett.105.182701} {\bibfield  {journal} {\bibinfo  {journal}
  {Phys. Rev. Lett.}\ }\textbf {\bibinfo {volume} {105}},\ \bibinfo {pages}
  {182701} (\bibinfo {year} {2010})}\BibitemShut {NoStop}%
\bibitem [{\citenamefont {Oganessian}\ \emph
  {et~al.}(2004{\natexlab{a}})\citenamefont {Oganessian}, \citenamefont
  {Utyonkoy}, \citenamefont {Lobanov}, \citenamefont {Abdullin}, \citenamefont
  {Polyakov}, \citenamefont {Shirokovsky}, \citenamefont {Tsyganov},
  \citenamefont {Gulbekian}, \citenamefont {Bogomolov}, \citenamefont
  {Mezentsev}, \citenamefont {Iliev}, \citenamefont {Subbotin}, \citenamefont
  {Sukhov}, \citenamefont {Voinov}, \citenamefont {Buklanov}, \citenamefont
  {Subotic}, \citenamefont {Zagrebaev}, \citenamefont {Itkis}, \citenamefont
  {Patin}, \citenamefont {Moody}, \citenamefont {Wild}, \citenamefont {Stoyer},
  \citenamefont {Stoyer}, \citenamefont {Shaughnessy}, \citenamefont
  {Kenneally},\ and\ \citenamefont {Lougheed}}]{PhysRevC.69.021601}%
  \BibitemOpen
  \bibfield  {author} {\bibinfo {author} {\bibfnamefont {Y.~T.}\ \bibnamefont
  {Oganessian}}, \bibinfo {author} {\bibfnamefont {V.~K.}\ \bibnamefont
  {Utyonkoy}}, \bibinfo {author} {\bibfnamefont {Y.~V.}\ \bibnamefont
  {Lobanov}}, \bibinfo {author} {\bibfnamefont {F.~S.}\ \bibnamefont
  {Abdullin}}, \bibinfo {author} {\bibfnamefont {A.~N.}\ \bibnamefont
  {Polyakov}}, \bibinfo {author} {\bibfnamefont {I.~V.}\ \bibnamefont
  {Shirokovsky}}, \bibinfo {author} {\bibfnamefont {Y.~S.}\ \bibnamefont
  {Tsyganov}}, \bibinfo {author} {\bibfnamefont {G.~G.}\ \bibnamefont
  {Gulbekian}}, \bibinfo {author} {\bibfnamefont {S.~L.}\ \bibnamefont
  {Bogomolov}}, \bibinfo {author} {\bibfnamefont {A.~N.}\ \bibnamefont
  {Mezentsev}}, \bibinfo {author} {\bibfnamefont {S.}~\bibnamefont {Iliev}},
  \bibinfo {author} {\bibfnamefont {V.~G.}\ \bibnamefont {Subbotin}}, \bibinfo
  {author} {\bibfnamefont {A.~M.}\ \bibnamefont {Sukhov}}, \bibinfo {author}
  {\bibfnamefont {A.~A.}\ \bibnamefont {Voinov}}, \bibinfo {author}
  {\bibfnamefont {G.~V.}\ \bibnamefont {Buklanov}}, \bibinfo {author}
  {\bibfnamefont {K.}~\bibnamefont {Subotic}}, \bibinfo {author} {\bibfnamefont
  {V.~I.}\ \bibnamefont {Zagrebaev}}, \bibinfo {author} {\bibfnamefont {M.~G.}\
  \bibnamefont {Itkis}}, \bibinfo {author} {\bibfnamefont {J.~B.}\ \bibnamefont
  {Patin}}, \bibinfo {author} {\bibfnamefont {K.~J.}\ \bibnamefont {Moody}},
  \bibinfo {author} {\bibfnamefont {J.~F.}\ \bibnamefont {Wild}}, \bibinfo
  {author} {\bibfnamefont {M.~A.}\ \bibnamefont {Stoyer}}, \bibinfo {author}
  {\bibfnamefont {N.~J.}\ \bibnamefont {Stoyer}}, \bibinfo {author}
  {\bibfnamefont {D.~A.}\ \bibnamefont {Shaughnessy}}, \bibinfo {author}
  {\bibfnamefont {J.~M.}\ \bibnamefont {Kenneally}}, \ and\ \bibinfo {author}
  {\bibfnamefont {R.~W.}\ \bibnamefont {Lougheed}},\ }\href {\doibase
  10.1103/PhysRevC.69.021601} {\bibfield  {journal} {\bibinfo  {journal} {Phys.
  Rev. C}\ }\textbf {\bibinfo {volume} {69}},\ \bibinfo {pages} {021601}
  (\bibinfo {year} {2004}{\natexlab{a}})}\BibitemShut {NoStop}%
\bibitem [{\citenamefont {Oganessian}\ \emph
  {et~al.}(2004{\natexlab{b}})\citenamefont {Oganessian}, \citenamefont
  {Utyonkov}, \citenamefont {Lobanov}, \citenamefont {Abdullin}, \citenamefont
  {Polyakov}, \citenamefont {Shirokovsky}, \citenamefont {Tsyganov},
  \citenamefont {Gulbekian}, \citenamefont {Bogomolov}, \citenamefont {Gikal},
  \citenamefont {Mezentsev}, \citenamefont {Iliev}, \citenamefont {Subbotin},
  \citenamefont {Sukhov}, \citenamefont {Voinov}, \citenamefont {Buklanov},
  \citenamefont {Subotic}, \citenamefont {Zagrebaev}, \citenamefont {Itkis},
  \citenamefont {Patin}, \citenamefont {Moody}, \citenamefont {Wild},
  \citenamefont {Stoyer}, \citenamefont {Stoyer}, \citenamefont {Shaughnessy},
  \citenamefont {Kenneally},\ and\ \citenamefont
  {Lougheed}}]{PhysRevC.69.054607}%
  \BibitemOpen
  \bibfield  {author} {\bibinfo {author} {\bibfnamefont {Y.~T.}\ \bibnamefont
  {Oganessian}}, \bibinfo {author} {\bibfnamefont {V.~K.}\ \bibnamefont
  {Utyonkov}}, \bibinfo {author} {\bibfnamefont {Y.~V.}\ \bibnamefont
  {Lobanov}}, \bibinfo {author} {\bibfnamefont {F.~S.}\ \bibnamefont
  {Abdullin}}, \bibinfo {author} {\bibfnamefont {A.~N.}\ \bibnamefont
  {Polyakov}}, \bibinfo {author} {\bibfnamefont {I.~V.}\ \bibnamefont
  {Shirokovsky}}, \bibinfo {author} {\bibfnamefont {Y.~S.}\ \bibnamefont
  {Tsyganov}}, \bibinfo {author} {\bibfnamefont {G.~G.}\ \bibnamefont
  {Gulbekian}}, \bibinfo {author} {\bibfnamefont {S.~L.}\ \bibnamefont
  {Bogomolov}}, \bibinfo {author} {\bibfnamefont {B.~N.}\ \bibnamefont
  {Gikal}}, \bibinfo {author} {\bibfnamefont {A.~N.}\ \bibnamefont
  {Mezentsev}}, \bibinfo {author} {\bibfnamefont {S.}~\bibnamefont {Iliev}},
  \bibinfo {author} {\bibfnamefont {V.~G.}\ \bibnamefont {Subbotin}}, \bibinfo
  {author} {\bibfnamefont {A.~M.}\ \bibnamefont {Sukhov}}, \bibinfo {author}
  {\bibfnamefont {A.~A.}\ \bibnamefont {Voinov}}, \bibinfo {author}
  {\bibfnamefont {G.~V.}\ \bibnamefont {Buklanov}}, \bibinfo {author}
  {\bibfnamefont {K.}~\bibnamefont {Subotic}}, \bibinfo {author} {\bibfnamefont
  {V.~I.}\ \bibnamefont {Zagrebaev}}, \bibinfo {author} {\bibfnamefont {M.~G.}\
  \bibnamefont {Itkis}}, \bibinfo {author} {\bibfnamefont {J.~B.}\ \bibnamefont
  {Patin}}, \bibinfo {author} {\bibfnamefont {K.~J.}\ \bibnamefont {Moody}},
  \bibinfo {author} {\bibfnamefont {J.~F.}\ \bibnamefont {Wild}}, \bibinfo
  {author} {\bibfnamefont {M.~A.}\ \bibnamefont {Stoyer}}, \bibinfo {author}
  {\bibfnamefont {N.~J.}\ \bibnamefont {Stoyer}}, \bibinfo {author}
  {\bibfnamefont {D.~A.}\ \bibnamefont {Shaughnessy}}, \bibinfo {author}
  {\bibfnamefont {J.~M.}\ \bibnamefont {Kenneally}}, \ and\ \bibinfo {author}
  {\bibfnamefont {R.~W.}\ \bibnamefont {Lougheed}},\ }\href {\doibase
  10.1103/PhysRevC.69.054607} {\bibfield  {journal} {\bibinfo  {journal} {Phys.
  Rev. C}\ }\textbf {\bibinfo {volume} {69}},\ \bibinfo {pages} {054607}
  (\bibinfo {year} {2004}{\natexlab{b}})}\BibitemShut {NoStop}%
\bibitem [{\citenamefont {Oganessian}\ \emph {et~al.}(2010)\citenamefont
  {Oganessian}, \citenamefont {Abdullin}, \citenamefont {Bailey}, \citenamefont
  {Benker}, \citenamefont {Bennett}, \citenamefont {Dmitriev}, \citenamefont
  {Ezold}, \citenamefont {Hamilton}, \citenamefont {Henderson}, \citenamefont
  {Itkis}, \citenamefont {Lobanov}, \citenamefont {Mezentsev}, \citenamefont
  {Moody}, \citenamefont {Nelson}, \citenamefont {Polyakov}, \citenamefont
  {Porter}, \citenamefont {Ramayya}, \citenamefont {Riley}, \citenamefont
  {Roberto}, \citenamefont {Ryabinin}, \citenamefont {Rykaczewski},
  \citenamefont {Sagaidak}, \citenamefont {Shaughnessy}, \citenamefont
  {Shirokovsky}, \citenamefont {Stoyer}, \citenamefont {Subbotin},
  \citenamefont {Sudowe}, \citenamefont {Sukhov}, \citenamefont {Tsyganov},
  \citenamefont {Utyonkov}, \citenamefont {Voinov}, \citenamefont {Vostokin},\
  and\ \citenamefont {Wilk}}]{PhysRevLett.104.142502}%
  \BibitemOpen
  \bibfield  {author} {\bibinfo {author} {\bibfnamefont {Y.~T.}\ \bibnamefont
  {Oganessian}}, \bibinfo {author} {\bibfnamefont {F.~S.}\ \bibnamefont
  {Abdullin}}, \bibinfo {author} {\bibfnamefont {P.~D.}\ \bibnamefont
  {Bailey}}, \bibinfo {author} {\bibfnamefont {D.~E.}\ \bibnamefont {Benker}},
  \bibinfo {author} {\bibfnamefont {M.~E.}\ \bibnamefont {Bennett}}, \bibinfo
  {author} {\bibfnamefont {S.~N.}\ \bibnamefont {Dmitriev}}, \bibinfo {author}
  {\bibfnamefont {J.~G.}\ \bibnamefont {Ezold}}, \bibinfo {author}
  {\bibfnamefont {J.~H.}\ \bibnamefont {Hamilton}}, \bibinfo {author}
  {\bibfnamefont {R.~A.}\ \bibnamefont {Henderson}}, \bibinfo {author}
  {\bibfnamefont {M.~G.}\ \bibnamefont {Itkis}}, \bibinfo {author}
  {\bibfnamefont {Y.~V.}\ \bibnamefont {Lobanov}}, \bibinfo {author}
  {\bibfnamefont {A.~N.}\ \bibnamefont {Mezentsev}}, \bibinfo {author}
  {\bibfnamefont {K.~J.}\ \bibnamefont {Moody}}, \bibinfo {author}
  {\bibfnamefont {S.~L.}\ \bibnamefont {Nelson}}, \bibinfo {author}
  {\bibfnamefont {A.~N.}\ \bibnamefont {Polyakov}}, \bibinfo {author}
  {\bibfnamefont {C.~E.}\ \bibnamefont {Porter}}, \bibinfo {author}
  {\bibfnamefont {A.~V.}\ \bibnamefont {Ramayya}}, \bibinfo {author}
  {\bibfnamefont {F.~D.}\ \bibnamefont {Riley}}, \bibinfo {author}
  {\bibfnamefont {J.~B.}\ \bibnamefont {Roberto}}, \bibinfo {author}
  {\bibfnamefont {M.~A.}\ \bibnamefont {Ryabinin}}, \bibinfo {author}
  {\bibfnamefont {K.~P.}\ \bibnamefont {Rykaczewski}}, \bibinfo {author}
  {\bibfnamefont {R.~N.}\ \bibnamefont {Sagaidak}}, \bibinfo {author}
  {\bibfnamefont {D.~A.}\ \bibnamefont {Shaughnessy}}, \bibinfo {author}
  {\bibfnamefont {I.~V.}\ \bibnamefont {Shirokovsky}}, \bibinfo {author}
  {\bibfnamefont {M.~A.}\ \bibnamefont {Stoyer}}, \bibinfo {author}
  {\bibfnamefont {V.~G.}\ \bibnamefont {Subbotin}}, \bibinfo {author}
  {\bibfnamefont {R.}~\bibnamefont {Sudowe}}, \bibinfo {author} {\bibfnamefont
  {A.~M.}\ \bibnamefont {Sukhov}}, \bibinfo {author} {\bibfnamefont {Y.~S.}\
  \bibnamefont {Tsyganov}}, \bibinfo {author} {\bibfnamefont {V.~K.}\
  \bibnamefont {Utyonkov}}, \bibinfo {author} {\bibfnamefont {A.~A.}\
  \bibnamefont {Voinov}}, \bibinfo {author} {\bibfnamefont {G.~K.}\
  \bibnamefont {Vostokin}}, \ and\ \bibinfo {author} {\bibfnamefont {P.~A.}\
  \bibnamefont {Wilk}},\ }\href {\doibase 10.1103/PhysRevLett.104.142502}
  {\bibfield  {journal} {\bibinfo  {journal} {Phys. Rev. Lett.}\ }\textbf
  {\bibinfo {volume} {104}},\ \bibinfo {pages} {142502} (\bibinfo {year}
  {2010})}\BibitemShut {NoStop}%
\bibitem [{\citenamefont {Oganessian}\ \emph {et~al.}(2006)\citenamefont
  {Oganessian}, \citenamefont {Utyonkov}, \citenamefont {Lobanov},
  \citenamefont {Abdullin}, \citenamefont {Polyakov}, \citenamefont {Sagaidak},
  \citenamefont {Shirokovsky}, \citenamefont {Tsyganov}, \citenamefont
  {Voinov}, \citenamefont {Gulbekian}, \citenamefont {Bogomolov}, \citenamefont
  {Gikal}, \citenamefont {Mezentsev}, \citenamefont {Iliev}, \citenamefont
  {Subbotin}, \citenamefont {Sukhov}, \citenamefont {Subotic}, \citenamefont
  {Zagrebaev}, \citenamefont {Vostokin}, \citenamefont {Itkis}, \citenamefont
  {Moody}, \citenamefont {Patin}, \citenamefont {Shaughnessy}, \citenamefont
  {Stoyer}, \citenamefont {Stoyer}, \citenamefont {Wilk}, \citenamefont
  {Kenneally}, \citenamefont {Landrum}, \citenamefont {Wild},\ and\
  \citenamefont {Lougheed}}]{PhysRevC.74.044602}%
  \BibitemOpen
  \bibfield  {author} {\bibinfo {author} {\bibfnamefont {Y.~T.}\ \bibnamefont
  {Oganessian}}, \bibinfo {author} {\bibfnamefont {V.~K.}\ \bibnamefont
  {Utyonkov}}, \bibinfo {author} {\bibfnamefont {Y.~V.}\ \bibnamefont
  {Lobanov}}, \bibinfo {author} {\bibfnamefont {F.~S.}\ \bibnamefont
  {Abdullin}}, \bibinfo {author} {\bibfnamefont {A.~N.}\ \bibnamefont
  {Polyakov}}, \bibinfo {author} {\bibfnamefont {R.~N.}\ \bibnamefont
  {Sagaidak}}, \bibinfo {author} {\bibfnamefont {I.~V.}\ \bibnamefont
  {Shirokovsky}}, \bibinfo {author} {\bibfnamefont {Y.~S.}\ \bibnamefont
  {Tsyganov}}, \bibinfo {author} {\bibfnamefont {A.~A.}\ \bibnamefont
  {Voinov}}, \bibinfo {author} {\bibfnamefont {G.~G.}\ \bibnamefont
  {Gulbekian}}, \bibinfo {author} {\bibfnamefont {S.~L.}\ \bibnamefont
  {Bogomolov}}, \bibinfo {author} {\bibfnamefont {B.~N.}\ \bibnamefont
  {Gikal}}, \bibinfo {author} {\bibfnamefont {A.~N.}\ \bibnamefont
  {Mezentsev}}, \bibinfo {author} {\bibfnamefont {S.}~\bibnamefont {Iliev}},
  \bibinfo {author} {\bibfnamefont {V.~G.}\ \bibnamefont {Subbotin}}, \bibinfo
  {author} {\bibfnamefont {A.~M.}\ \bibnamefont {Sukhov}}, \bibinfo {author}
  {\bibfnamefont {K.}~\bibnamefont {Subotic}}, \bibinfo {author} {\bibfnamefont
  {V.~I.}\ \bibnamefont {Zagrebaev}}, \bibinfo {author} {\bibfnamefont {G.~K.}\
  \bibnamefont {Vostokin}}, \bibinfo {author} {\bibfnamefont {M.~G.}\
  \bibnamefont {Itkis}}, \bibinfo {author} {\bibfnamefont {K.~J.}\ \bibnamefont
  {Moody}}, \bibinfo {author} {\bibfnamefont {J.~B.}\ \bibnamefont {Patin}},
  \bibinfo {author} {\bibfnamefont {D.~A.}\ \bibnamefont {Shaughnessy}},
  \bibinfo {author} {\bibfnamefont {M.~A.}\ \bibnamefont {Stoyer}}, \bibinfo
  {author} {\bibfnamefont {N.~J.}\ \bibnamefont {Stoyer}}, \bibinfo {author}
  {\bibfnamefont {P.~A.}\ \bibnamefont {Wilk}}, \bibinfo {author}
  {\bibfnamefont {J.~M.}\ \bibnamefont {Kenneally}}, \bibinfo {author}
  {\bibfnamefont {J.~H.}\ \bibnamefont {Landrum}}, \bibinfo {author}
  {\bibfnamefont {J.~F.}\ \bibnamefont {Wild}}, \ and\ \bibinfo {author}
  {\bibfnamefont {R.~W.}\ \bibnamefont {Lougheed}},\ }\href {\doibase
  10.1103/PhysRevC.74.044602} {\bibfield  {journal} {\bibinfo  {journal} {Phys.
  Rev. C}\ }\textbf {\bibinfo {volume} {74}},\ \bibinfo {pages} {044602}
  (\bibinfo {year} {2006})}\BibitemShut {NoStop}%
\bibitem [{\citenamefont {Oganesyan}\ \emph {et~al.}(1975)\citenamefont
  {Oganesyan}, \citenamefont {Tretyakov}, \citenamefont {Ilinov}, \citenamefont
  {Demin}, \citenamefont {Pleve}, \citenamefont {Tretyakova}, \citenamefont
  {Plotko}, \citenamefont {Ivanov}, \citenamefont {Danilov}, \citenamefont
  {Korotkin},\ and\ \citenamefont {Flerov}}]{1975Og}%
  \BibitemOpen
  \bibfield  {author} {\bibinfo {author} {\bibfnamefont {Y.~T.}\ \bibnamefont
  {Oganesyan}}, \bibinfo {author} {\bibfnamefont {Y.~P.}\ \bibnamefont
  {Tretyakov}}, \bibinfo {author} {\bibfnamefont {A.~S.}\ \bibnamefont
  {Ilinov}}, \bibinfo {author} {\bibfnamefont {A.~G.}\ \bibnamefont {Demin}},
  \bibinfo {author} {\bibfnamefont {A.~A.}\ \bibnamefont {Pleve}}, \bibinfo
  {author} {\bibfnamefont {S.~P.}\ \bibnamefont {Tretyakova}}, \bibinfo
  {author} {\bibfnamefont {V.~M.}\ \bibnamefont {Plotko}}, \bibinfo {author}
  {\bibfnamefont {M.~P.}\ \bibnamefont {Ivanov}}, \bibinfo {author}
  {\bibfnamefont {N.~A.}\ \bibnamefont {Danilov}}, \bibinfo {author}
  {\bibfnamefont {Y.~S.}\ \bibnamefont {Korotkin}}, \ and\ \bibinfo {author}
  {\bibfnamefont {G.~N.}\ \bibnamefont {Flerov}},\ }\href@noop {} {\bibfield
  {journal} {\bibinfo  {journal} {JETP Lett. (USSR)}\ }\textbf {\bibinfo
  {volume} {20}},\ \bibinfo {pages} {265} (\bibinfo {year} {1975})}\BibitemShut
  {NoStop}%
\bibitem [{\citenamefont {{M{\"u}nzenberg}}\ \emph {et~al.}(1981)\citenamefont
  {{M{\"u}nzenberg}}, \citenamefont {{Hofmann}}, \citenamefont
  {{He{\ss}berger}}, \citenamefont {{Reisdorf}}, \citenamefont {{Schmidt}},
  \citenamefont {{Schneider}}, \citenamefont {{Armbruster}}, \citenamefont
  {{Sahm}},\ and\ \citenamefont {{Thuma}}}]{1981ZPhyA300107M}%
  \BibitemOpen
  \bibfield  {author} {\bibinfo {author} {\bibfnamefont {G.}~\bibnamefont
  {{M{\"u}nzenberg}}}, \bibinfo {author} {\bibfnamefont {S.}~\bibnamefont
  {{Hofmann}}}, \bibinfo {author} {\bibfnamefont {F.~P.}\ \bibnamefont
  {{He{\ss}berger}}}, \bibinfo {author} {\bibfnamefont {W.}~\bibnamefont
  {{Reisdorf}}}, \bibinfo {author} {\bibfnamefont {K.~H.}\ \bibnamefont
  {{Schmidt}}}, \bibinfo {author} {\bibfnamefont {J.~H.~R.}\ \bibnamefont
  {{Schneider}}}, \bibinfo {author} {\bibfnamefont {P.}~\bibnamefont
  {{Armbruster}}}, \bibinfo {author} {\bibfnamefont {C.~C.}\ \bibnamefont
  {{Sahm}}}, \ and\ \bibinfo {author} {\bibfnamefont {B.}~\bibnamefont
  {{Thuma}}},\ }\href {\doibase 10.1007/BF01412623} {\bibfield  {journal}
  {\bibinfo  {journal} {Eur. Phys. J. A}\ }\textbf {\bibinfo {volume} {300}},\
  \bibinfo {pages} {107} (\bibinfo {year} {1981})}\BibitemShut {NoStop}%
\bibitem [{\citenamefont {Münzenberg}\ \emph {et~al.}(1984)\citenamefont
  {Münzenberg}, \citenamefont {Armbruster}, \citenamefont {Folger},
  \citenamefont {Heßberger}, \citenamefont {Hofmann}, \citenamefont {Keller},
  \citenamefont {Poppensieker}, \citenamefont {Reisdorf}, \citenamefont
  {Schmidt}, \citenamefont {Schött}, \citenamefont {Leino},\ and\
  \citenamefont {Hingmann}}]{article84mu}%
  \BibitemOpen
  \bibfield  {author} {\bibinfo {author} {\bibfnamefont {G.}~\bibnamefont
  {Münzenberg}}, \bibinfo {author} {\bibfnamefont {P.}~\bibnamefont
  {Armbruster}}, \bibinfo {author} {\bibfnamefont {H.}~\bibnamefont {Folger}},
  \bibinfo {author} {\bibfnamefont {P.}~\bibnamefont {Heßberger}}, \bibinfo
  {author} {\bibfnamefont {S.}~\bibnamefont {Hofmann}}, \bibinfo {author}
  {\bibfnamefont {J.}~\bibnamefont {Keller}}, \bibinfo {author} {\bibfnamefont
  {K.}~\bibnamefont {Poppensieker}}, \bibinfo {author} {\bibfnamefont
  {W.}~\bibnamefont {Reisdorf}}, \bibinfo {author} {\bibfnamefont
  {K.}~\bibnamefont {Schmidt}}, \bibinfo {author} {\bibfnamefont
  {H.}~\bibnamefont {Schött}}, \bibinfo {author} {\bibfnamefont
  {M.}~\bibnamefont {Leino}}, \ and\ \bibinfo {author} {\bibfnamefont
  {R.}~\bibnamefont {Hingmann}},\ }\href {\doibase 10.1007/BF01421260}
  {\bibfield  {journal} {\bibinfo  {journal} {Eur. Phys. J. A}\ }\textbf
  {\bibinfo {volume} {317}},\ \bibinfo {pages} {235} (\bibinfo {year}
  {1984})}\BibitemShut {NoStop}%
\bibitem [{\citenamefont {Hofmann}\ \emph
  {et~al.}(1995{\natexlab{a}})\citenamefont {Hofmann}, \citenamefont {Ninov},
  \citenamefont {Heßberger}, \citenamefont {Armbruster}, \citenamefont
  {Folger}, \citenamefont {Münzenberg}, \citenamefont {Schött}, \citenamefont
  {Popeko}, \citenamefont {Yeremin}, \citenamefont {Andreyev}, \citenamefont
  {Saro}, \citenamefont {Janik},\ and\ \citenamefont {Leino}}]{articlehof95}%
  \BibitemOpen
  \bibfield  {author} {\bibinfo {author} {\bibfnamefont {S.}~\bibnamefont
  {Hofmann}}, \bibinfo {author} {\bibfnamefont {V.}~\bibnamefont {Ninov}},
  \bibinfo {author} {\bibfnamefont {F.}~\bibnamefont {Heßberger}}, \bibinfo
  {author} {\bibfnamefont {P.}~\bibnamefont {Armbruster}}, \bibinfo {author}
  {\bibfnamefont {H.}~\bibnamefont {Folger}}, \bibinfo {author} {\bibfnamefont
  {G.}~\bibnamefont {Münzenberg}}, \bibinfo {author} {\bibfnamefont
  {H.}~\bibnamefont {Schött}}, \bibinfo {author} {\bibfnamefont
  {A.}~\bibnamefont {Popeko}}, \bibinfo {author} {\bibfnamefont
  {A.}~\bibnamefont {Yeremin}}, \bibinfo {author} {\bibfnamefont
  {A.}~\bibnamefont {Andreyev}}, \bibinfo {author} {\bibfnamefont
  {S.}~\bibnamefont {Saro}}, \bibinfo {author} {\bibfnamefont {R.}~\bibnamefont
  {Janik}}, \ and\ \bibinfo {author} {\bibfnamefont {M.}~\bibnamefont
  {Leino}},\ }\href {\doibase 10.1007/BF01291181} {\bibfield  {journal}
  {\bibinfo  {journal} {Eur. Phys. J. A}\ }\textbf {\bibinfo {volume} {350}},\
  \bibinfo {pages} {277} (\bibinfo {year} {1995}{\natexlab{a}})}\BibitemShut
  {NoStop}%
\bibitem [{\citenamefont {Hofmann}\ \emph
  {et~al.}(1995{\natexlab{b}})\citenamefont {Hofmann}, \citenamefont {Ninov},
  \citenamefont {Heßberger}, \citenamefont {Armbruster}, \citenamefont
  {Folger}, \citenamefont {Münzenberg}, \citenamefont {Schött}, \citenamefont
  {Popeko}, \citenamefont {Yeremin}, \citenamefont {Andreyev}, \citenamefont
  {Saro}, \citenamefont {Janik},\ and\ \citenamefont {Leino}}]{articlehof111}%
  \BibitemOpen
  \bibfield  {author} {\bibinfo {author} {\bibfnamefont {S.}~\bibnamefont
  {Hofmann}}, \bibinfo {author} {\bibfnamefont {V.}~\bibnamefont {Ninov}},
  \bibinfo {author} {\bibfnamefont {F.}~\bibnamefont {Heßberger}}, \bibinfo
  {author} {\bibfnamefont {P.}~\bibnamefont {Armbruster}}, \bibinfo {author}
  {\bibfnamefont {H.}~\bibnamefont {Folger}}, \bibinfo {author} {\bibfnamefont
  {G.}~\bibnamefont {Münzenberg}}, \bibinfo {author} {\bibfnamefont
  {H.}~\bibnamefont {Schött}}, \bibinfo {author} {\bibfnamefont
  {A.}~\bibnamefont {Popeko}}, \bibinfo {author} {\bibfnamefont
  {A.}~\bibnamefont {Yeremin}}, \bibinfo {author} {\bibfnamefont
  {A.}~\bibnamefont {Andreyev}}, \bibinfo {author} {\bibfnamefont
  {S.}~\bibnamefont {Saro}}, \bibinfo {author} {\bibfnamefont {R.}~\bibnamefont
  {Janik}}, \ and\ \bibinfo {author} {\bibfnamefont {M.}~\bibnamefont
  {Leino}},\ }\href {\doibase 10.1007/BF01291182} {\bibfield  {journal}
  {\bibinfo  {journal} {Eur. Phys. J. A}\ }\textbf {\bibinfo {volume} {350}},\
  \bibinfo {pages} {281} (\bibinfo {year} {1995}{\natexlab{b}})}\BibitemShut
  {NoStop}%
\bibitem [{\citenamefont {Hofmann}\ \emph {et~al.}(1996)\citenamefont
  {Hofmann}, \citenamefont {Ninov}, \citenamefont {Heßberger}, \citenamefont
  {Armbruster}, \citenamefont {Folger}, \citenamefont {Munzenberg},
  \citenamefont {Schött}, \citenamefont {Popeko}, \citenamefont {Yeremin},
  \citenamefont {Saro}, \citenamefont {Janik},\ and\ \citenamefont
  {Leino}}]{Hof96112}%
  \BibitemOpen
  \bibfield  {author} {\bibinfo {author} {\bibfnamefont {S.}~\bibnamefont
  {Hofmann}}, \bibinfo {author} {\bibfnamefont {V.}~\bibnamefont {Ninov}},
  \bibinfo {author} {\bibfnamefont {F.~P.}\ \bibnamefont {Heßberger}},
  \bibinfo {author} {\bibfnamefont {P.}~\bibnamefont {Armbruster}}, \bibinfo
  {author} {\bibfnamefont {H.}~\bibnamefont {Folger}}, \bibinfo {author}
  {\bibfnamefont {G.}~\bibnamefont {Munzenberg}}, \bibinfo {author}
  {\bibfnamefont {H.~J.}\ \bibnamefont {Schött}}, \bibinfo {author}
  {\bibfnamefont {A.~G.}\ \bibnamefont {Popeko}}, \bibinfo {author}
  {\bibfnamefont {A.~V.}\ \bibnamefont {Yeremin}}, \bibinfo {author}
  {\bibfnamefont {S.}~\bibnamefont {Saro}}, \bibinfo {author} {\bibfnamefont
  {R.}~\bibnamefont {Janik}}, \ and\ \bibinfo {author} {\bibfnamefont
  {M.}~\bibnamefont {Leino}},\ }\href {\doibase 10.1007/BF02769517} {\bibfield
  {journal} {\bibinfo  {journal} {Eur. Phys. J. A}\ }\textbf {\bibinfo {volume}
  {354}},\ \bibinfo {pages} {229} (\bibinfo {year} {1996})}\BibitemShut
  {NoStop}%
\bibitem [{\citenamefont {Morita}\ \emph {et~al.}(2004)\citenamefont {Morita},
  \citenamefont {Morimoto}, \citenamefont {Kaji}, \citenamefont {Akiyama},
  \citenamefont {Goto}, \citenamefont {Haba}, \citenamefont {Ideguchi},
  \citenamefont {Kanungo}, \citenamefont {Katori}, \citenamefont {Koura},
  \citenamefont {Kudo}, \citenamefont {Ohnishi}, \citenamefont {Ozawa},
  \citenamefont {Suda}, \citenamefont {Sueki}, \citenamefont {Xu},
  \citenamefont {Yamaguchi}, \citenamefont {Yoneda}, \citenamefont {Yoshida},\
  and\ \citenamefont {Zhao}}]{doi:10.1143/JPSJ.73.2593}%
  \BibitemOpen
  \bibfield  {author} {\bibinfo {author} {\bibfnamefont {K.}~\bibnamefont
  {Morita}}, \bibinfo {author} {\bibfnamefont {K.}~\bibnamefont {Morimoto}},
  \bibinfo {author} {\bibfnamefont {D.}~\bibnamefont {Kaji}}, \bibinfo {author}
  {\bibfnamefont {T.}~\bibnamefont {Akiyama}}, \bibinfo {author} {\bibfnamefont
  {S.-i.}\ \bibnamefont {Goto}}, \bibinfo {author} {\bibfnamefont
  {H.}~\bibnamefont {Haba}}, \bibinfo {author} {\bibfnamefont {E.}~\bibnamefont
  {Ideguchi}}, \bibinfo {author} {\bibfnamefont {R.}~\bibnamefont {Kanungo}},
  \bibinfo {author} {\bibfnamefont {K.}~\bibnamefont {Katori}}, \bibinfo
  {author} {\bibfnamefont {H.}~\bibnamefont {Koura}}, \bibinfo {author}
  {\bibfnamefont {H.}~\bibnamefont {Kudo}}, \bibinfo {author} {\bibfnamefont
  {T.}~\bibnamefont {Ohnishi}}, \bibinfo {author} {\bibfnamefont
  {A.}~\bibnamefont {Ozawa}}, \bibinfo {author} {\bibfnamefont
  {T.}~\bibnamefont {Suda}}, \bibinfo {author} {\bibfnamefont {K.}~\bibnamefont
  {Sueki}}, \bibinfo {author} {\bibfnamefont {H.}~\bibnamefont {Xu}}, \bibinfo
  {author} {\bibfnamefont {T.}~\bibnamefont {Yamaguchi}}, \bibinfo {author}
  {\bibfnamefont {A.}~\bibnamefont {Yoneda}}, \bibinfo {author} {\bibfnamefont
  {A.}~\bibnamefont {Yoshida}}, \ and\ \bibinfo {author} {\bibfnamefont
  {Y.}~\bibnamefont {Zhao}},\ }\href {\doibase 10.1143/JPSJ.73.2593} {\bibfield
   {journal} {\bibinfo  {journal} {J. Phys. Soc. J.}\ }\textbf {\bibinfo
  {volume} {73}},\ \bibinfo {pages} {2593} (\bibinfo {year} {2004})},\ \Eprint
  {http://arxiv.org/abs/https://doi.org/10.1143/JPSJ.73.2593}
  {https://doi.org/10.1143/JPSJ.73.2593} \BibitemShut {NoStop}%
\bibitem [{\citenamefont {Gan}\ \emph {et~al.}(2001)\citenamefont {Gan},
  \citenamefont {Qin}, \citenamefont {Fan}, \citenamefont {Lei}, \citenamefont
  {Xu}, \citenamefont {He}, \citenamefont {Liu}, \citenamefont {Wu},
  \citenamefont {Guo}, \citenamefont {Zhou}, \citenamefont {Yuan},\ and\
  \citenamefont {Jin}}]{epjagan01}%
  \BibitemOpen
  \bibfield  {author} {\bibinfo {author} {\bibfnamefont {Z.}~\bibnamefont
  {Gan}}, \bibinfo {author} {\bibfnamefont {Z.}~\bibnamefont {Qin}}, \bibinfo
  {author} {\bibfnamefont {H.}~\bibnamefont {Fan}}, \bibinfo {author}
  {\bibfnamefont {X.}~\bibnamefont {Lei}}, \bibinfo {author} {\bibfnamefont
  {Y.}~\bibnamefont {Xu}}, \bibinfo {author} {\bibfnamefont {J.}~\bibnamefont
  {He}}, \bibinfo {author} {\bibfnamefont {H.}~\bibnamefont {Liu}}, \bibinfo
  {author} {\bibfnamefont {X.}~\bibnamefont {Wu}}, \bibinfo {author}
  {\bibfnamefont {J.-s.}\ \bibnamefont {Guo}}, \bibinfo {author} {\bibfnamefont
  {X.}~\bibnamefont {Zhou}}, \bibinfo {author} {\bibfnamefont {S.}~\bibnamefont
  {Yuan}}, \ and\ \bibinfo {author} {\bibfnamefont {G.}~\bibnamefont {Jin}},\
  }\href {\doibase 10.1007/s100500170140} {\bibfield  {journal} {\bibinfo
  {journal} {Eur. Phys. J. A}\ }\textbf {\bibinfo {volume} {10}},\ \bibinfo
  {pages} {21} (\bibinfo {year} {2001})}\BibitemShut {NoStop}%
\bibitem [{\citenamefont {Qin}\ \emph {et~al.}(2006)\citenamefont {Qin},
  \citenamefont {Wu}, \citenamefont {Ding}, \citenamefont {Wu}, \citenamefont
  {Huang}, \citenamefont {Lei}, \citenamefont {Xu}, \citenamefont {Yuan},
  \citenamefont {Guo}, \citenamefont {Yang}, \citenamefont {Gan}, \citenamefont
  {Fan}, \citenamefont {Guo}, \citenamefont {Xu},\ and\ \citenamefont
  {Xiao}}]{06-4-8npr06}%
  \BibitemOpen
  \bibfield  {author} {\bibinfo {author} {\bibfnamefont {Z.}~\bibnamefont
  {Qin}}, \bibinfo {author} {\bibfnamefont {X.~L.}\ \bibnamefont {Wu}},
  \bibinfo {author} {\bibfnamefont {H.~J.}\ \bibnamefont {Ding}}, \bibinfo
  {author} {\bibfnamefont {W.~S.}\ \bibnamefont {Wu}}, \bibinfo {author}
  {\bibfnamefont {W.~X.}\ \bibnamefont {Huang}}, \bibinfo {author}
  {\bibfnamefont {X.~G.}\ \bibnamefont {Lei}}, \bibinfo {author} {\bibfnamefont
  {Y.~B.}\ \bibnamefont {Xu}}, \bibinfo {author} {\bibfnamefont {X.~H.}\
  \bibnamefont {Yuan}}, \bibinfo {author} {\bibfnamefont {B.}~\bibnamefont
  {Guo}}, \bibinfo {author} {\bibfnamefont {W.~F.}\ \bibnamefont {Yang}},
  \bibinfo {author} {\bibfnamefont {Z.~G.}\ \bibnamefont {Gan}}, \bibinfo
  {author} {\bibfnamefont {H.~M.}\ \bibnamefont {Fan}}, \bibinfo {author}
  {\bibfnamefont {J.~S.}\ \bibnamefont {Guo}}, \bibinfo {author} {\bibfnamefont
  {H.~S.}\ \bibnamefont {Xu}}, \ and\ \bibinfo {author} {\bibfnamefont {G.-Q.}\
  \bibnamefont {Xiao}},\ }\href {\doibase 10.11804/NuclPhysRev.23.04.404}
  {\bibfield  {journal} {\bibinfo  {journal} {Nucl. Phys. Rev.}\ }\textbf
  {\bibinfo {volume} {23}},\ \bibinfo {pages} {404} (\bibinfo {year}
  {2006})}\BibitemShut {NoStop}%
\bibitem [{\citenamefont {Zhang}\ \emph {et~al.}(2012)\citenamefont {Zhang},
  \citenamefont {Gan}, \citenamefont {Ma}, \citenamefont {Huang}, \citenamefont
  {Huang}, \citenamefont {Wu}, \citenamefont {Jia}, \citenamefont {Li},
  \citenamefont {Yu}, \citenamefont {Ren}, \citenamefont {Zhou}, \citenamefont
  {Zhang}, \citenamefont {Zhou}, \citenamefont {Xu}, \citenamefont {Zhang},
  \citenamefont {Xiao},\ and\ \citenamefont {Zhan}}]{cpl12zhang}%
  \BibitemOpen
  \bibfield  {author} {\bibinfo {author} {\bibfnamefont {Z.-Y.}\ \bibnamefont
  {Zhang}}, \bibinfo {author} {\bibfnamefont {Z.-G.}\ \bibnamefont {Gan}},
  \bibinfo {author} {\bibfnamefont {L.}~\bibnamefont {Ma}}, \bibinfo {author}
  {\bibfnamefont {M.-H.}\ \bibnamefont {Huang}}, \bibinfo {author}
  {\bibfnamefont {T.-H.}\ \bibnamefont {Huang}}, \bibinfo {author}
  {\bibfnamefont {X.-L.}\ \bibnamefont {Wu}}, \bibinfo {author} {\bibfnamefont
  {G.-B.}\ \bibnamefont {Jia}}, \bibinfo {author} {\bibfnamefont {G.-S.}\
  \bibnamefont {Li}}, \bibinfo {author} {\bibfnamefont {L.}~\bibnamefont {Yu}},
  \bibinfo {author} {\bibfnamefont {Z.-Z.}\ \bibnamefont {Ren}}, \bibinfo
  {author} {\bibfnamefont {S.-G.}\ \bibnamefont {Zhou}}, \bibinfo {author}
  {\bibfnamefont {Y.-H.}\ \bibnamefont {Zhang}}, \bibinfo {author}
  {\bibfnamefont {X.-H.}\ \bibnamefont {Zhou}}, \bibinfo {author}
  {\bibfnamefont {H.-S.}\ \bibnamefont {Xu}}, \bibinfo {author} {\bibfnamefont
  {H.-Q.}\ \bibnamefont {Zhang}}, \bibinfo {author} {\bibfnamefont {G.-Q.}\
  \bibnamefont {Xiao}}, \ and\ \bibinfo {author} {\bibfnamefont {W.-L.}\
  \bibnamefont {Zhan}},\ }\href {\doibase 10.1088/0256-307x/29/1/012502}
  {\bibfield  {journal} {\bibinfo  {journal} {Chin. Phys. Lett.}\ }\textbf
  {\bibinfo {volume} {29}},\ \bibinfo {pages} {012502} (\bibinfo {year}
  {2012})}\BibitemShut {NoStop}%
\bibitem [{\citenamefont {Niu}\ \emph {et~al.}(2021{\natexlab{a}})\citenamefont
  {Niu}, \citenamefont {Chen},\ and\ \citenamefont {Feng}}]{nst2021niu}%
  \BibitemOpen
  \bibfield  {author} {\bibinfo {author} {\bibfnamefont {F.}~\bibnamefont
  {Niu}}, \bibinfo {author} {\bibfnamefont {P.-H.}\ \bibnamefont {Chen}}, \
  and\ \bibinfo {author} {\bibfnamefont {Z.-Q.}\ \bibnamefont {Feng}},\ }\href
  {\doibase 10.1007/s41365-021-00946-3} {\bibfield  {journal} {\bibinfo
  {journal} {Nucl. Sci. Tec.}\ }\textbf {\bibinfo {volume} {32}},\ \bibinfo
  {pages} {103} (\bibinfo {year} {2021}{\natexlab{a}})}\BibitemShut {NoStop}%
\bibitem [{\citenamefont {Guo}\ \emph {et~al.}(2018)\citenamefont {Guo},
  \citenamefont {Simenel}, \citenamefont {Shi},\ and\ \citenamefont
  {Yu}}]{GUO2018401}%
  \BibitemOpen
  \bibfield  {author} {\bibinfo {author} {\bibfnamefont {L.}~\bibnamefont
  {Guo}}, \bibinfo {author} {\bibfnamefont {C.}~\bibnamefont {Simenel}},
  \bibinfo {author} {\bibfnamefont {L.}~\bibnamefont {Shi}}, \ and\ \bibinfo
  {author} {\bibfnamefont {C.}~\bibnamefont {Yu}},\ }\href {\doibase
  https://doi.org/10.1016/j.physletb.2018.05.066} {\bibfield  {journal}
  {\bibinfo  {journal} {Phys. Lett. B}\ }\textbf {\bibinfo {volume} {782}},\
  \bibinfo {pages} {401} (\bibinfo {year} {2018})}\BibitemShut {NoStop}%
\bibitem [{\citenamefont {Sekizawa}(2019)}]{10.3389/fphy.2019.00020}%
  \BibitemOpen
  \bibfield  {author} {\bibinfo {author} {\bibfnamefont {K.}~\bibnamefont
  {Sekizawa}},\ }\href {\doibase 10.3389/fphy.2019.00020} {\bibfield  {journal}
  {\bibinfo  {journal} {Fron. Phys.}\ }\textbf {\bibinfo {volume} {7}}
  (\bibinfo {year} {2019}),\ 10.3389/fphy.2019.00020}\BibitemShut {NoStop}%
\bibitem [{\citenamefont {Maruhn}\ \emph {et~al.}(2014)\citenamefont {Maruhn},
  \citenamefont {Reinhard}, \citenamefont {Stevenson},\ and\ \citenamefont
  {Umar}}]{MARUHN20142195}%
  \BibitemOpen
  \bibfield  {author} {\bibinfo {author} {\bibfnamefont {J.}~\bibnamefont
  {Maruhn}}, \bibinfo {author} {\bibfnamefont {P.-G.}\ \bibnamefont
  {Reinhard}}, \bibinfo {author} {\bibfnamefont {P.}~\bibnamefont {Stevenson}},
  \ and\ \bibinfo {author} {\bibfnamefont {A.}~\bibnamefont {Umar}},\ }\href
  {\doibase https://doi.org/10.1016/j.cpc.2014.04.008} {\bibfield  {journal}
  {\bibinfo  {journal} {Com. Phys. Comm.}\ }\textbf {\bibinfo {volume} {185}},\
  \bibinfo {pages} {2195} (\bibinfo {year} {2014})}\BibitemShut {NoStop}%
\bibitem [{\citenamefont {Wang}\ \emph {et~al.}(2002)\citenamefont {Wang},
  \citenamefont {Li},\ and\ \citenamefont {Wu}}]{PhysRevC.65.064608}%
  \BibitemOpen
  \bibfield  {author} {\bibinfo {author} {\bibfnamefont {N.}~\bibnamefont
  {Wang}}, \bibinfo {author} {\bibfnamefont {Z.}~\bibnamefont {Li}}, \ and\
  \bibinfo {author} {\bibfnamefont {X.}~\bibnamefont {Wu}},\ }\href {\doibase
  10.1103/PhysRevC.65.064608} {\bibfield  {journal} {\bibinfo  {journal} {Phys.
  Rev. C}\ }\textbf {\bibinfo {volume} {65}},\ \bibinfo {pages} {064608}
  (\bibinfo {year} {2002})}\BibitemShut {NoStop}%
\bibitem [{\citenamefont {Jiang}\ \emph {et~al.}(2013)\citenamefont {Jiang},
  \citenamefont {Yan},\ and\ \citenamefont {Maruhn}}]{PhysRevC.88.044611}%
  \BibitemOpen
  \bibfield  {author} {\bibinfo {author} {\bibfnamefont {X.}~\bibnamefont
  {Jiang}}, \bibinfo {author} {\bibfnamefont {S.}~\bibnamefont {Yan}}, \ and\
  \bibinfo {author} {\bibfnamefont {J.~A.}\ \bibnamefont {Maruhn}},\ }\href
  {\doibase 10.1103/PhysRevC.88.044611} {\bibfield  {journal} {\bibinfo
  {journal} {Phys. Rev. C}\ }\textbf {\bibinfo {volume} {88}},\ \bibinfo
  {pages} {044611} (\bibinfo {year} {2013})}\BibitemShut {NoStop}%
\bibitem [{\citenamefont {Zhao}\ \emph {et~al.}(2013)\citenamefont {Zhao},
  \citenamefont {Li}, \citenamefont {Wu},\ and\ \citenamefont
  {Zhang}}]{PhysRevC.88.044605}%
  \BibitemOpen
  \bibfield  {author} {\bibinfo {author} {\bibfnamefont {K.}~\bibnamefont
  {Zhao}}, \bibinfo {author} {\bibfnamefont {Z.}~\bibnamefont {Li}}, \bibinfo
  {author} {\bibfnamefont {X.}~\bibnamefont {Wu}}, \ and\ \bibinfo {author}
  {\bibfnamefont {Y.}~\bibnamefont {Zhang}},\ }\href {\doibase
  10.1103/PhysRevC.88.044605} {\bibfield  {journal} {\bibinfo  {journal} {Phys.
  Rev. C}\ }\textbf {\bibinfo {volume} {88}},\ \bibinfo {pages} {044605}
  (\bibinfo {year} {2013})}\BibitemShut {NoStop}%
\bibitem [{\citenamefont {Yanez}\ and\ \citenamefont
  {Loveland}(2015)}]{Grazing}%
  \BibitemOpen
  \bibfield  {author} {\bibinfo {author} {\bibfnamefont {R.}~\bibnamefont
  {Yanez}}\ and\ \bibinfo {author} {\bibfnamefont {W.}~\bibnamefont
  {Loveland}},\ }\href {\doibase 10.1103/PhysRevC.91.044608} {\bibfield
  {journal} {\bibinfo  {journal} {Phys. Rev. C}\ }\textbf {\bibinfo {volume}
  {91}},\ \bibinfo {pages} {044608} (\bibinfo {year} {2015})}\BibitemShut
  {NoStop}%
\bibitem [{\citenamefont {Feng}\ \emph {et~al.}(2006)\citenamefont {Feng},
  \citenamefont {Jin}, \citenamefont {Fu},\ and\ \citenamefont
  {Li}}]{FENG200650}%
  \BibitemOpen
  \bibfield  {author} {\bibinfo {author} {\bibfnamefont {Z.-Q.}\ \bibnamefont
  {Feng}}, \bibinfo {author} {\bibfnamefont {G.-M.}\ \bibnamefont {Jin}},
  \bibinfo {author} {\bibfnamefont {F.}~\bibnamefont {Fu}}, \ and\ \bibinfo
  {author} {\bibfnamefont {J.-Q.}\ \bibnamefont {Li}},\ }\href {\doibase
  https://doi.org/10.1016/j.nuclphysa.2006.03.002} {\bibfield  {journal}
  {\bibinfo  {journal} {Nucl. Phys. A}\ }\textbf {\bibinfo {volume} {771}},\
  \bibinfo {pages} {50} (\bibinfo {year} {2006})}\BibitemShut {NoStop}%
\bibitem [{\citenamefont {Bao}\ \emph {et~al.}(2015)\citenamefont {Bao},
  \citenamefont {Gao}, \citenamefont {Li},\ and\ \citenamefont
  {Zhang}}]{PhysRevC.91.011603}%
  \BibitemOpen
  \bibfield  {author} {\bibinfo {author} {\bibfnamefont {X.~J.}\ \bibnamefont
  {Bao}}, \bibinfo {author} {\bibfnamefont {Y.}~\bibnamefont {Gao}}, \bibinfo
  {author} {\bibfnamefont {J.~Q.}\ \bibnamefont {Li}}, \ and\ \bibinfo {author}
  {\bibfnamefont {H.~F.}\ \bibnamefont {Zhang}},\ }\href {\doibase
  10.1103/PhysRevC.91.011603} {\bibfield  {journal} {\bibinfo  {journal} {Phys.
  Rev. C}\ }\textbf {\bibinfo {volume} {91}},\ \bibinfo {pages} {011603}
  (\bibinfo {year} {2015})}\BibitemShut {NoStop}%
\bibitem [{\citenamefont {Zhu}\ \emph {et~al.}(2014{\natexlab{a}})\citenamefont
  {Zhu}, \citenamefont {Xie},\ and\ \citenamefont
  {Zhang}}]{PhysRevC.89.024615}%
  \BibitemOpen
  \bibfield  {author} {\bibinfo {author} {\bibfnamefont {L.}~\bibnamefont
  {Zhu}}, \bibinfo {author} {\bibfnamefont {W.-J.}\ \bibnamefont {Xie}}, \ and\
  \bibinfo {author} {\bibfnamefont {F.-S.}\ \bibnamefont {Zhang}},\ }\href
  {\doibase 10.1103/PhysRevC.89.024615} {\bibfield  {journal} {\bibinfo
  {journal} {Phys. Rev. C}\ }\textbf {\bibinfo {volume} {89}},\ \bibinfo
  {pages} {024615} (\bibinfo {year} {2014}{\natexlab{a}})}\BibitemShut
  {NoStop}%
\bibitem [{\citenamefont {Adamyan}\ \emph {et~al.}(2020)\citenamefont
  {Adamyan}, \citenamefont {Antonenko}, \citenamefont {Diaz-Torres},\ and\
  \citenamefont {Heinz}}]{epja20gga}%
  \BibitemOpen
  \bibfield  {author} {\bibinfo {author} {\bibfnamefont {G.}~\bibnamefont
  {Adamyan}}, \bibinfo {author} {\bibfnamefont {N.}~\bibnamefont {Antonenko}},
  \bibinfo {author} {\bibfnamefont {A.}~\bibnamefont {Diaz-Torres}}, \ and\
  \bibinfo {author} {\bibfnamefont {S.}~\bibnamefont {Heinz}},\ }\href
  {\doibase 10.1140/epja/s10050-020-00046-7} {\bibfield  {journal} {\bibinfo
  {journal} {Eur. Phys. J. A}\ }\textbf {\bibinfo {volume} {56}},\ \bibinfo
  {pages} {47} (\bibinfo {year} {2020})}\BibitemShut {NoStop}%
\bibitem [{\citenamefont {Feng}\ \emph
  {et~al.}(2007{\natexlab{a}})\citenamefont {Feng}, \citenamefont {Jin},
  \citenamefont {Huang}, \citenamefont {Gan}, \citenamefont {Wang},\ and\
  \citenamefont {Li}}]{07fengcpl}%
  \BibitemOpen
  \bibfield  {author} {\bibinfo {author} {\bibfnamefont {Z.-Q.}\ \bibnamefont
  {Feng}}, \bibinfo {author} {\bibfnamefont {G.-M.}\ \bibnamefont {Jin}},
  \bibinfo {author} {\bibfnamefont {M.-H.}\ \bibnamefont {Huang}}, \bibinfo
  {author} {\bibfnamefont {Z.-G.}\ \bibnamefont {Gan}}, \bibinfo {author}
  {\bibfnamefont {N.}~\bibnamefont {Wang}}, \ and\ \bibinfo {author}
  {\bibfnamefont {J.-Q.}\ \bibnamefont {Li}},\ }\href {\doibase
  10.1088/0256-307x/24/9/024} {\bibfield  {journal} {\bibinfo  {journal} {Chin.
  Phys. Lett.}\ }\textbf {\bibinfo {volume} {24}},\ \bibinfo {pages} {2551}
  (\bibinfo {year} {2007}{\natexlab{a}})}\BibitemShut {NoStop}%
\bibitem [{\citenamefont {Möller}\ \emph {et~al.}(2016)\citenamefont
  {Möller}, \citenamefont {Sierk}, \citenamefont {Ichikawa},\ and\
  \citenamefont {Sagawa}}]{MOLLER20161}%
  \BibitemOpen
  \bibfield  {author} {\bibinfo {author} {\bibfnamefont {P.}~\bibnamefont
  {Möller}}, \bibinfo {author} {\bibfnamefont {A.}~\bibnamefont {Sierk}},
  \bibinfo {author} {\bibfnamefont {T.}~\bibnamefont {Ichikawa}}, \ and\
  \bibinfo {author} {\bibfnamefont {H.}~\bibnamefont {Sagawa}},\ }\href
  {\doibase https://doi.org/10.1016/j.adt.2015.10.002} {\bibfield  {journal}
  {\bibinfo  {journal} {Atomic Data and Nuclear Data Tables}\ }\textbf
  {\bibinfo {volume} {109-110}},\ \bibinfo {pages} {1} (\bibinfo {year}
  {2016})}\BibitemShut {NoStop}%
\bibitem [{\citenamefont {Koura}\ \emph {et~al.}(2005)\citenamefont {Koura},
  \citenamefont {Tachibana}, \citenamefont {Uno},\ and\ \citenamefont
  {Yamada}}]{10.1143/PTP.113.305}%
  \BibitemOpen
  \bibfield  {author} {\bibinfo {author} {\bibfnamefont {H.}~\bibnamefont
  {Koura}}, \bibinfo {author} {\bibfnamefont {T.}~\bibnamefont {Tachibana}},
  \bibinfo {author} {\bibfnamefont {M.}~\bibnamefont {Uno}}, \ and\ \bibinfo
  {author} {\bibfnamefont {M.}~\bibnamefont {Yamada}},\ }\href {\doibase
  10.1143/PTP.113.305} {\bibfield  {journal} {\bibinfo  {journal} {Progress of
  Theoretical Physics}\ }\textbf {\bibinfo {volume} {113}},\ \bibinfo {pages}
  {305} (\bibinfo {year} {2005})},\ \Eprint
  {http://arxiv.org/abs/https://academic.oup.com/ptp/article-pdf/113/2/305/5192381/113-2-305.pdf}
  {https://academic.oup.com/ptp/article-pdf/113/2/305/5192381/113-2-305.pdf}
  \BibitemShut {NoStop}%
\bibitem [{\citenamefont {Wang}\ \emph
  {et~al.}(2014{\natexlab{a}})\citenamefont {Wang}, \citenamefont {Liu},
  \citenamefont {Wu},\ and\ \citenamefont {Meng}}]{WANG2014215}%
  \BibitemOpen
  \bibfield  {author} {\bibinfo {author} {\bibfnamefont {N.}~\bibnamefont
  {Wang}}, \bibinfo {author} {\bibfnamefont {M.}~\bibnamefont {Liu}}, \bibinfo
  {author} {\bibfnamefont {X.}~\bibnamefont {Wu}}, \ and\ \bibinfo {author}
  {\bibfnamefont {J.}~\bibnamefont {Meng}},\ }\href {\doibase
  https://doi.org/10.1016/j.physletb.2014.05.049} {\bibfield  {journal}
  {\bibinfo  {journal} {Physics Letters B}\ }\textbf {\bibinfo {volume}
  {734}},\ \bibinfo {pages} {215} (\bibinfo {year}
  {2014}{\natexlab{a}})}\BibitemShut {NoStop}%
\bibitem [{\citenamefont {Myers}\ and\ \citenamefont
  {Swiatecki}(1966)}]{MYERS19661}%
  \BibitemOpen
  \bibfield  {author} {\bibinfo {author} {\bibfnamefont {W.~D.}\ \bibnamefont
  {Myers}}\ and\ \bibinfo {author} {\bibfnamefont {W.~J.}\ \bibnamefont
  {Swiatecki}},\ }\href {\doibase https://doi.org/10.1016/0029-5582(66)90639-0}
  {\bibfield  {journal} {\bibinfo  {journal} {Nucl. Phys.}\ }\textbf {\bibinfo
  {volume} {81}},\ \bibinfo {pages} {1} (\bibinfo {year} {1966})}\BibitemShut
  {NoStop}%
\bibitem [{\citenamefont {{El Bassem}}\ and\ \citenamefont
  {Oulne}(2017)}]{ELBASSEM201722}%
  \BibitemOpen
  \bibfield  {author} {\bibinfo {author} {\bibfnamefont {Y.}~\bibnamefont {{El
  Bassem}}}\ and\ \bibinfo {author} {\bibfnamefont {M.}~\bibnamefont {Oulne}},\
  }\href {\doibase https://doi.org/10.1016/j.nuclphysa.2016.07.005} {\bibfield
  {journal} {\bibinfo  {journal} {Nuclear Physics A}\ }\textbf {\bibinfo
  {volume} {957}},\ \bibinfo {pages} {22} (\bibinfo {year} {2017})}\BibitemShut
  {NoStop}%
\bibitem [{\citenamefont {Feng}\ \emph
  {et~al.}(2010{\natexlab{a}})\citenamefont {Feng}, \citenamefont {Jin},\ and\
  \citenamefont {Li}}]{FENG201082}%
  \BibitemOpen
  \bibfield  {author} {\bibinfo {author} {\bibfnamefont {Z.-Q.}\ \bibnamefont
  {Feng}}, \bibinfo {author} {\bibfnamefont {G.-M.}\ \bibnamefont {Jin}}, \
  and\ \bibinfo {author} {\bibfnamefont {J.-Q.}\ \bibnamefont {Li}},\ }\href
  {\doibase https://doi.org/10.1016/j.nuclphysa.2010.01.244} {\bibfield
  {journal} {\bibinfo  {journal} {Nucl. Phys. A}\ }\textbf {\bibinfo {volume}
  {836}},\ \bibinfo {pages} {82} (\bibinfo {year}
  {2010}{\natexlab{a}})}\BibitemShut {NoStop}%
\bibitem [{\citenamefont {Chen}\ \emph {et~al.}(2023)\citenamefont {Chen},
  \citenamefont {Wu}, \citenamefont {Yang}, \citenamefont {Zeng},\ and\
  \citenamefont {Feng}}]{Chen2023}%
  \BibitemOpen
  \bibfield  {author} {\bibinfo {author} {\bibfnamefont {P.-H.}\ \bibnamefont
  {Chen}}, \bibinfo {author} {\bibfnamefont {H.}~\bibnamefont {Wu}}, \bibinfo
  {author} {\bibfnamefont {Z.-X.}\ \bibnamefont {Yang}}, \bibinfo {author}
  {\bibfnamefont {X.-H.}\ \bibnamefont {Zeng}}, \ and\ \bibinfo {author}
  {\bibfnamefont {Z.-Q.}\ \bibnamefont {Feng}},\ }\href {\doibase
  10.1007/s41365-022-01157-0} {\bibfield  {journal} {\bibinfo  {journal}
  {Nuclear Science and Techniques}\ }\textbf {\bibinfo {volume} {34}},\
  \bibinfo {pages} {7} (\bibinfo {year} {2023})}\BibitemShut {NoStop}%
\bibitem [{\citenamefont {Feng}\ \emph
  {et~al.}(2009{\natexlab{a}})\citenamefont {Feng}, \citenamefont {Jin},\ and\
  \citenamefont {Li}}]{PhysRevC.80.057601}%
  \BibitemOpen
  \bibfield  {author} {\bibinfo {author} {\bibfnamefont {Z.-Q.}\ \bibnamefont
  {Feng}}, \bibinfo {author} {\bibfnamefont {G.-M.}\ \bibnamefont {Jin}}, \
  and\ \bibinfo {author} {\bibfnamefont {J.-Q.}\ \bibnamefont {Li}},\ }\href
  {\doibase 10.1103/PhysRevC.80.057601} {\bibfield  {journal} {\bibinfo
  {journal} {Phys. Rev. C}\ }\textbf {\bibinfo {volume} {80}},\ \bibinfo
  {pages} {057601} (\bibinfo {year} {2009}{\natexlab{a}})}\BibitemShut
  {NoStop}%
\bibitem [{\citenamefont {Feng}\ \emph
  {et~al.}(2007{\natexlab{b}})\citenamefont {Feng}, \citenamefont {Jin},
  \citenamefont {Li},\ and\ \citenamefont {Scheid}}]{PhysRevC.76.044606}%
  \BibitemOpen
  \bibfield  {author} {\bibinfo {author} {\bibfnamefont {Z.-Q.}\ \bibnamefont
  {Feng}}, \bibinfo {author} {\bibfnamefont {G.-M.}\ \bibnamefont {Jin}},
  \bibinfo {author} {\bibfnamefont {J.-Q.}\ \bibnamefont {Li}}, \ and\ \bibinfo
  {author} {\bibfnamefont {W.}~\bibnamefont {Scheid}},\ }\href {\doibase
  10.1103/PhysRevC.76.044606} {\bibfield  {journal} {\bibinfo  {journal} {Phys.
  Rev. C}\ }\textbf {\bibinfo {volume} {76}},\ \bibinfo {pages} {044606}
  (\bibinfo {year} {2007}{\natexlab{b}})}\BibitemShut {NoStop}%
\bibitem [{\citenamefont {Hill}\ and\ \citenamefont
  {Wheeler}(1953)}]{PhysRev.89.1102}%
  \BibitemOpen
  \bibfield  {author} {\bibinfo {author} {\bibfnamefont {D.~L.}\ \bibnamefont
  {Hill}}\ and\ \bibinfo {author} {\bibfnamefont {J.~A.}\ \bibnamefont
  {Wheeler}},\ }\href {\doibase 10.1103/PhysRev.89.1102} {\bibfield  {journal}
  {\bibinfo  {journal} {Phys. Rev.}\ }\textbf {\bibinfo {volume} {89}},\
  \bibinfo {pages} {1102} (\bibinfo {year} {1953})}\BibitemShut {NoStop}%
\bibitem [{\citenamefont {Zagrebaev}\ \emph {et~al.}(2001)\citenamefont
  {Zagrebaev}, \citenamefont {Aritomo}, \citenamefont {Itkis}, \citenamefont
  {Oganessian},\ and\ \citenamefont {Ohta}}]{PhysRevC.65.014607}%
  \BibitemOpen
  \bibfield  {author} {\bibinfo {author} {\bibfnamefont {V.~I.}\ \bibnamefont
  {Zagrebaev}}, \bibinfo {author} {\bibfnamefont {Y.}~\bibnamefont {Aritomo}},
  \bibinfo {author} {\bibfnamefont {M.~G.}\ \bibnamefont {Itkis}}, \bibinfo
  {author} {\bibfnamefont {Y.~T.}\ \bibnamefont {Oganessian}}, \ and\ \bibinfo
  {author} {\bibfnamefont {M.}~\bibnamefont {Ohta}},\ }\href {\doibase
  10.1103/PhysRevC.65.014607} {\bibfield  {journal} {\bibinfo  {journal} {Phys.
  Rev. C}\ }\textbf {\bibinfo {volume} {65}},\ \bibinfo {pages} {014607}
  (\bibinfo {year} {2001})}\BibitemShut {NoStop}%
\bibitem [{\citenamefont {Cheng}\ \emph {et~al.}(2022)\citenamefont {Cheng},
  \citenamefont {Pu}, \citenamefont {Wang}, \citenamefont {Guo},\ and\
  \citenamefont {Ma}}]{Cheng2022}%
  \BibitemOpen
  \bibfield  {author} {\bibinfo {author} {\bibfnamefont {K.-X.}\ \bibnamefont
  {Cheng}}, \bibinfo {author} {\bibfnamefont {J.}~\bibnamefont {Pu}}, \bibinfo
  {author} {\bibfnamefont {Y.-T.}\ \bibnamefont {Wang}}, \bibinfo {author}
  {\bibfnamefont {Y.-F.}\ \bibnamefont {Guo}}, \ and\ \bibinfo {author}
  {\bibfnamefont {C.-W.}\ \bibnamefont {Ma}},\ }\href {\doibase
  10.1007/s41365-022-01114-x} {\bibfield  {journal} {\bibinfo  {journal}
  {Nuclear Science and Techniques}\ }\textbf {\bibinfo {volume} {33}},\
  \bibinfo {pages} {132} (\bibinfo {year} {2022})}\BibitemShut {NoStop}%
\bibitem [{\citenamefont {Adamian}\ \emph {et~al.}(1996)\citenamefont
  {Adamian}, \citenamefont {Antonenko}, \citenamefont {Jolos}, \citenamefont
  {Ivanova},\ and\ \citenamefont {Melnikova}}]{J.Mod.Phys.E5191(1996)}%
  \BibitemOpen
  \bibfield  {author} {\bibinfo {author} {\bibfnamefont {G.~G.}\ \bibnamefont
  {Adamian}}, \bibinfo {author} {\bibfnamefont {N.~V.}\ \bibnamefont
  {Antonenko}}, \bibinfo {author} {\bibfnamefont {R.~V.}\ \bibnamefont
  {Jolos}}, \bibinfo {author} {\bibfnamefont {S.~P.}\ \bibnamefont {Ivanova}},
  \ and\ \bibinfo {author} {\bibfnamefont {O.~I.}\ \bibnamefont {Melnikova}},\
  }\href {https://doi.org/10.1142/S0218301396000098} {\bibfield  {journal}
  {\bibinfo  {journal} {Inter. J. Mod. Phys. E}\ }\textbf {\bibinfo {volume}
  {05}},\ \bibinfo {pages} {191216} (\bibinfo {year} {1996})}\BibitemShut
  {NoStop}%
\bibitem [{\citenamefont {Wong}(1973)}]{PhysRevLett.31.766}%
  \BibitemOpen
  \bibfield  {author} {\bibinfo {author} {\bibfnamefont {C.~Y.}\ \bibnamefont
  {Wong}},\ }\href {\doibase 10.1103/PhysRevLett.31.766} {\bibfield  {journal}
  {\bibinfo  {journal} {Phys. Rev. Lett.}\ }\textbf {\bibinfo {volume} {31}},\
  \bibinfo {pages} {766} (\bibinfo {year} {1973})}\BibitemShut {NoStop}%
\bibitem [{\citenamefont {Feng}\ \emph
  {et~al.}(2009{\natexlab{b}})\citenamefont {Feng}, \citenamefont {Jin},
  \citenamefont {Li},\ and\ \citenamefont {Scheid}}]{FENG200933}%
  \BibitemOpen
  \bibfield  {author} {\bibinfo {author} {\bibfnamefont {Z.-Q.}\ \bibnamefont
  {Feng}}, \bibinfo {author} {\bibfnamefont {G.-M.}\ \bibnamefont {Jin}},
  \bibinfo {author} {\bibfnamefont {J.-Q.}\ \bibnamefont {Li}}, \ and\ \bibinfo
  {author} {\bibfnamefont {W.}~\bibnamefont {Scheid}},\ }\href {\doibase
  https://doi.org/10.1016/j.nuclphysa.2008.11.003} {\bibfield  {journal}
  {\bibinfo  {journal} {Nucl. Phys. A}\ }\textbf {\bibinfo {volume} {816}},\
  \bibinfo {pages} {33} (\bibinfo {year} {2009}{\natexlab{b}})}\BibitemShut
  {NoStop}%
\bibitem [{\citenamefont {Li}\ and\ \citenamefont
  {Wolschin}(1983)}]{PhysRevC.27.590}%
  \BibitemOpen
  \bibfield  {author} {\bibinfo {author} {\bibfnamefont {J.~Q.}\ \bibnamefont
  {Li}}\ and\ \bibinfo {author} {\bibfnamefont {G.}~\bibnamefont {Wolschin}},\
  }\href {\doibase 10.1103/PhysRevC.27.590} {\bibfield  {journal} {\bibinfo
  {journal} {Phys. Rev. C}\ }\textbf {\bibinfo {volume} {27}},\ \bibinfo
  {pages} {590} (\bibinfo {year} {1983})}\BibitemShut {NoStop}%
\bibitem [{\citenamefont {Li}\ \emph {et~al.}(1981)\citenamefont {Li},
  \citenamefont {Tang},\ and\ \citenamefont {Wolschin}}]{LI1981107}%
  \BibitemOpen
  \bibfield  {author} {\bibinfo {author} {\bibfnamefont {J.}~\bibnamefont
  {Li}}, \bibinfo {author} {\bibfnamefont {X.}~\bibnamefont {Tang}}, \ and\
  \bibinfo {author} {\bibfnamefont {G.}~\bibnamefont {Wolschin}},\ }\href
  {\doibase https://doi.org/10.1016/0370-2693(81)91000-5} {\bibfield  {journal}
  {\bibinfo  {journal} {Phys. Lett. B}\ }\textbf {\bibinfo {volume} {105}},\
  \bibinfo {pages} {107} (\bibinfo {year} {1981})}\BibitemShut {NoStop}%
\bibitem [{\citenamefont {Adamian}\ \emph {et~al.}(2003)\citenamefont
  {Adamian}, \citenamefont {Antonenko},\ and\ \citenamefont
  {Scheid}}]{PhysRevC.68.034601}%
  \BibitemOpen
  \bibfield  {author} {\bibinfo {author} {\bibfnamefont {G.~G.}\ \bibnamefont
  {Adamian}}, \bibinfo {author} {\bibfnamefont {N.~V.}\ \bibnamefont
  {Antonenko}}, \ and\ \bibinfo {author} {\bibfnamefont {W.}~\bibnamefont
  {Scheid}},\ }\href {\doibase 10.1103/PhysRevC.68.034601} {\bibfield
  {journal} {\bibinfo  {journal} {Phys. Rev. C}\ }\textbf {\bibinfo {volume}
  {68}},\ \bibinfo {pages} {034601} (\bibinfo {year} {2003})}\BibitemShut
  {NoStop}%
\bibitem [{\citenamefont {Grang\'e}\ \emph {et~al.}(1983)\citenamefont
  {Grang\'e}, \citenamefont {Jun-Qing},\ and\ \citenamefont
  {Weidenm\"uller}}]{PhysRevC.27.2063}%
  \BibitemOpen
  \bibfield  {author} {\bibinfo {author} {\bibfnamefont {P.}~\bibnamefont
  {Grang\'e}}, \bibinfo {author} {\bibfnamefont {L.}~\bibnamefont {Jun-Qing}},
  \ and\ \bibinfo {author} {\bibfnamefont {H.~A.}\ \bibnamefont
  {Weidenm\"uller}},\ }\href {\doibase 10.1103/PhysRevC.27.2063} {\bibfield
  {journal} {\bibinfo  {journal} {Phys. Rev. C}\ }\textbf {\bibinfo {volume}
  {27}},\ \bibinfo {pages} {2063} (\bibinfo {year} {1983})}\BibitemShut
  {NoStop}%
\bibitem [{\citenamefont {Chen}\ \emph {et~al.}(2016)\citenamefont {Chen},
  \citenamefont {Feng}, \citenamefont {Li},\ and\ \citenamefont
  {Zhang}}]{Chen_2016}%
  \BibitemOpen
  \bibfield  {author} {\bibinfo {author} {\bibfnamefont {P.-H.}\ \bibnamefont
  {Chen}}, \bibinfo {author} {\bibfnamefont {Z.-Q.}\ \bibnamefont {Feng}},
  \bibinfo {author} {\bibfnamefont {J.-Q.}\ \bibnamefont {Li}}, \ and\ \bibinfo
  {author} {\bibfnamefont {H.-F.}\ \bibnamefont {Zhang}},\ }\href {\doibase
  10.1088/1674-1137/40/9/091002} {\bibfield  {journal} {\bibinfo  {journal}
  {Chin. Phys. C}\ }\textbf {\bibinfo {volume} {40}},\ \bibinfo {pages}
  {091002} (\bibinfo {year} {2016})}\BibitemShut {NoStop}%
\bibitem [{\citenamefont {Xin}\ \emph {et~al.}(2021)\citenamefont {Xin},
  \citenamefont {Ma}, \citenamefont {Deng}, \citenamefont {Zhao},\ and\
  \citenamefont {Zhang}}]{Xin2021}%
  \BibitemOpen
  \bibfield  {author} {\bibinfo {author} {\bibfnamefont {Y.-Q.}\ \bibnamefont
  {Xin}}, \bibinfo {author} {\bibfnamefont {N.-N.}\ \bibnamefont {Ma}},
  \bibinfo {author} {\bibfnamefont {J.-G.}\ \bibnamefont {Deng}}, \bibinfo
  {author} {\bibfnamefont {T.-L.}\ \bibnamefont {Zhao}}, \ and\ \bibinfo
  {author} {\bibfnamefont {H.-F.}\ \bibnamefont {Zhang}},\ }\href {\doibase
  10.1007/s41365-021-00899-7} {\bibfield  {journal} {\bibinfo  {journal}
  {Nuclear Science and Techniques}\ }\textbf {\bibinfo {volume} {32}},\
  \bibinfo {pages} {55} (\bibinfo {year} {2021})}\BibitemShut {NoStop}%
\bibitem [{\citenamefont {Zubov}\ \emph {et~al.}(2003)\citenamefont {Zubov},
  \citenamefont {Adamian}, \citenamefont {Antonenko}, \citenamefont {Ivanova},\
  and\ \citenamefont {Scheid}}]{PhysRevC.68.014616}%
  \BibitemOpen
  \bibfield  {author} {\bibinfo {author} {\bibfnamefont {A.~S.}\ \bibnamefont
  {Zubov}}, \bibinfo {author} {\bibfnamefont {G.~G.}\ \bibnamefont {Adamian}},
  \bibinfo {author} {\bibfnamefont {N.~V.}\ \bibnamefont {Antonenko}}, \bibinfo
  {author} {\bibfnamefont {S.~P.}\ \bibnamefont {Ivanova}}, \ and\ \bibinfo
  {author} {\bibfnamefont {W.}~\bibnamefont {Scheid}},\ }\href {\doibase
  10.1103/PhysRevC.68.014616} {\bibfield  {journal} {\bibinfo  {journal} {Phys.
  Rev. C}\ }\textbf {\bibinfo {volume} {68}},\ \bibinfo {pages} {014616}
  (\bibinfo {year} {2003})}\BibitemShut {NoStop}%
\bibitem [{\citenamefont {Zubov}\ \emph {et~al.}(2005)\citenamefont {Zubov},
  \citenamefont {Adamyan}, \citenamefont {Antonenko}, \citenamefont {Ivanova},\
  and\ \citenamefont {Scheid}}]{artza05}%
  \BibitemOpen
  \bibfield  {author} {\bibinfo {author} {\bibfnamefont {A.}~\bibnamefont
  {Zubov}}, \bibinfo {author} {\bibfnamefont {G.~G.}\ \bibnamefont {Adamyan}},
  \bibinfo {author} {\bibfnamefont {N.}~\bibnamefont {Antonenko}}, \bibinfo
  {author} {\bibfnamefont {S.}~\bibnamefont {Ivanova}}, \ and\ \bibinfo
  {author} {\bibfnamefont {W.}~\bibnamefont {Scheid}},\ }\href {\doibase
  10.1140/epja/i2004-10089-5} {\bibfield  {journal} {\bibinfo  {journal} {Eur.
  Phys. J. A}\ }\textbf {\bibinfo {volume} {23}},\ \bibinfo {pages} {249}
  (\bibinfo {year} {2005})}\BibitemShut {NoStop}%
\bibitem [{\citenamefont {Moller}\ \emph {et~al.}(1995)\citenamefont {Moller},
  \citenamefont {Nix}, \citenamefont {Myers},\ and\ \citenamefont
  {Swiatecki}}]{Moller_1995}%
  \BibitemOpen
  \bibfield  {author} {\bibinfo {author} {\bibfnamefont {P.}~\bibnamefont
  {Moller}}, \bibinfo {author} {\bibfnamefont {J.}~\bibnamefont {Nix}},
  \bibinfo {author} {\bibfnamefont {W.}~\bibnamefont {Myers}}, \ and\ \bibinfo
  {author} {\bibfnamefont {W.}~\bibnamefont {Swiatecki}},\ }\href {\doibase
  10.1006/adnd.1995.1002} {\bibfield  {journal} {\bibinfo  {journal} {Atom Dat
  Nucl. Dat Tab.}\ }\textbf {\bibinfo {volume} {59}},\ \bibinfo {pages} {185}
  (\bibinfo {year} {1995})}\BibitemShut {NoStop}%
\bibitem [{\citenamefont {FENG}\ \emph {et~al.}(2011)\citenamefont {FENG},
  \citenamefont {JIN},\ and\ \citenamefont {LI}}]{2011fennpr}%
  \BibitemOpen
  \bibfield  {author} {\bibinfo {author} {\bibfnamefont {Z.-Q.}\ \bibnamefont
  {FENG}}, \bibinfo {author} {\bibfnamefont {G.-M.}\ \bibnamefont {JIN}}, \
  and\ \bibinfo {author} {\bibfnamefont {J.-Q.}\ \bibnamefont {LI}},\ }\href
  {\doibase 10.11804/NuclPhysRev.28.01.001} {\bibfield  {journal} {\bibinfo
  {journal} {Nucl. Phys. Rev.}\ }\textbf {\bibinfo {volume} {28}},\ \bibinfo
  {pages} {1} (\bibinfo {year} {2011})}\BibitemShut {NoStop}%
\bibitem [{\citenamefont {Wang}\ \emph {et~al.}(2008)\citenamefont {Wang},
  \citenamefont {Li},\ and\ \citenamefont {Zhao}}]{PhysRevC.78.054607}%
  \BibitemOpen
  \bibfield  {author} {\bibinfo {author} {\bibfnamefont {N.}~\bibnamefont
  {Wang}}, \bibinfo {author} {\bibfnamefont {J.-q.}\ \bibnamefont {Li}}, \ and\
  \bibinfo {author} {\bibfnamefont {E.-g.}\ \bibnamefont {Zhao}},\ }\href
  {\doibase 10.1103/PhysRevC.78.054607} {\bibfield  {journal} {\bibinfo
  {journal} {Phys. Rev. C}\ }\textbf {\bibinfo {volume} {78}},\ \bibinfo
  {pages} {054607} (\bibinfo {year} {2008})}\BibitemShut {NoStop}%
\bibitem [{\citenamefont {Feng}\ \emph
  {et~al.}(2010{\natexlab{b}})\citenamefont {Feng}, \citenamefont {Jin},
  \citenamefont {Li},\ and\ \citenamefont {Scheid}}]{FENG2010384c}%
  \BibitemOpen
  \bibfield  {author} {\bibinfo {author} {\bibfnamefont {Z.-Q.}\ \bibnamefont
  {Feng}}, \bibinfo {author} {\bibfnamefont {G.-M.}\ \bibnamefont {Jin}},
  \bibinfo {author} {\bibfnamefont {J.-Q.}\ \bibnamefont {Li}}, \ and\ \bibinfo
  {author} {\bibfnamefont {W.}~\bibnamefont {Scheid}},\ }\href {\doibase
  https://doi.org/10.1016/j.nuclphysa.2010.01.046} {\bibfield  {journal}
  {\bibinfo  {journal} {Nucl. Phys. A}\ }\textbf {\bibinfo {volume} {834}},\
  \bibinfo {pages} {384c} (\bibinfo {year} {2010}{\natexlab{b}})},\ \bibinfo
  {note} {the 10th International Conference on Nucleus-Nucleus Collisions
  (NN2009)}\BibitemShut {NoStop}%
\bibitem [{\citenamefont {Zhu}\ \emph {et~al.}(2014{\natexlab{b}})\citenamefont
  {Zhu}, \citenamefont {Feng}, \citenamefont {Li},\ and\ \citenamefont
  {Zhang}}]{PhysRevC.90.014612}%
  \BibitemOpen
  \bibfield  {author} {\bibinfo {author} {\bibfnamefont {L.}~\bibnamefont
  {Zhu}}, \bibinfo {author} {\bibfnamefont {Z.-Q.}\ \bibnamefont {Feng}},
  \bibinfo {author} {\bibfnamefont {C.}~\bibnamefont {Li}}, \ and\ \bibinfo
  {author} {\bibfnamefont {F.-S.}\ \bibnamefont {Zhang}},\ }\href {\doibase
  10.1103/PhysRevC.90.014612} {\bibfield  {journal} {\bibinfo  {journal} {Phys.
  Rev. C}\ }\textbf {\bibinfo {volume} {90}},\ \bibinfo {pages} {014612}
  (\bibinfo {year} {2014}{\natexlab{b}})}\BibitemShut {NoStop}%
\bibitem [{\citenamefont {Bao}(2019)}]{PhysRevC.100.011601}%
  \BibitemOpen
  \bibfield  {author} {\bibinfo {author} {\bibfnamefont {X.~J.}\ \bibnamefont
  {Bao}},\ }\href {\doibase 10.1103/PhysRevC.100.011601} {\bibfield  {journal}
  {\bibinfo  {journal} {Phys. Rev. C}\ }\textbf {\bibinfo {volume} {100}},\
  \bibinfo {pages} {011601} (\bibinfo {year} {2019})}\BibitemShut {NoStop}%
\bibitem [{\citenamefont {ZHAO}\ \emph {et~al.}(2008)\citenamefont {ZHAO},
  \citenamefont {WANG}, \citenamefont {FENG}, \citenamefont {LI}, \citenamefont
  {ZHOU},\ and\ \citenamefont {SCHEID}}]{doi:10.1142/S021830130801091X}%
  \BibitemOpen
  \bibfield  {author} {\bibinfo {author} {\bibfnamefont {E.~G.}\ \bibnamefont
  {ZHAO}}, \bibinfo {author} {\bibfnamefont {N.}~\bibnamefont {WANG}}, \bibinfo
  {author} {\bibfnamefont {Z.~Q.}\ \bibnamefont {FENG}}, \bibinfo {author}
  {\bibfnamefont {J.~Q.}\ \bibnamefont {LI}}, \bibinfo {author} {\bibfnamefont
  {S.~G.}\ \bibnamefont {ZHOU}}, \ and\ \bibinfo {author} {\bibfnamefont
  {W.}~\bibnamefont {SCHEID}},\ }\href {\doibase 10.1142/S021830130801091X}
  {\bibfield  {journal} {\bibinfo  {journal} {Int. J. Mod. Phys. E}\ }\textbf
  {\bibinfo {volume} {17}},\ \bibinfo {pages} {1937} (\bibinfo {year}
  {2008})},\ \Eprint
  {http://arxiv.org/abs/https://doi.org/10.1142/S021830130801091X}
  {https://doi.org/10.1142/S021830130801091X} \BibitemShut {NoStop}%
\bibitem [{\citenamefont {Wang}\ \emph
  {et~al.}(2014{\natexlab{b}})\citenamefont {Wang}, \citenamefont {Zhao},\ and\
  \citenamefont {Scheid}}]{PhysRevC.89.037601}%
  \BibitemOpen
  \bibfield  {author} {\bibinfo {author} {\bibfnamefont {N.}~\bibnamefont
  {Wang}}, \bibinfo {author} {\bibfnamefont {E.-G.}\ \bibnamefont {Zhao}}, \
  and\ \bibinfo {author} {\bibfnamefont {W.}~\bibnamefont {Scheid}},\ }\href
  {\doibase 10.1103/PhysRevC.89.037601} {\bibfield  {journal} {\bibinfo
  {journal} {Phys. Rev. C}\ }\textbf {\bibinfo {volume} {89}},\ \bibinfo
  {pages} {037601} (\bibinfo {year} {2014}{\natexlab{b}})}\BibitemShut
  {NoStop}%
\bibitem [{\citenamefont {Wang}\ \emph {et~al.}(2012)\citenamefont {Wang},
  \citenamefont {Zhao}, \citenamefont {Scheid},\ and\ \citenamefont
  {Zhou}}]{PhysRevC.85.041601}%
  \BibitemOpen
  \bibfield  {author} {\bibinfo {author} {\bibfnamefont {N.}~\bibnamefont
  {Wang}}, \bibinfo {author} {\bibfnamefont {E.-G.}\ \bibnamefont {Zhao}},
  \bibinfo {author} {\bibfnamefont {W.}~\bibnamefont {Scheid}}, \ and\ \bibinfo
  {author} {\bibfnamefont {S.-G.}\ \bibnamefont {Zhou}},\ }\href {\doibase
  10.1103/PhysRevC.85.041601} {\bibfield  {journal} {\bibinfo  {journal} {Phys.
  Rev. C}\ }\textbf {\bibinfo {volume} {85}},\ \bibinfo {pages} {041601}
  (\bibinfo {year} {2012})}\BibitemShut {NoStop}%
\bibitem [{\citenamefont {Zhao-Qing}\ \emph {et~al.}(2009)\citenamefont
  {Zhao-Qing}, \citenamefont {Gen-Ming}, \citenamefont {Jun-Qing},\ and\
  \citenamefont {Scheid}}]{FZQ2009}%
  \BibitemOpen
  \bibfield  {author} {\bibinfo {author} {\bibfnamefont {F.}~\bibnamefont
  {Zhao-Qing}}, \bibinfo {author} {\bibfnamefont {J.}~\bibnamefont {Gen-Ming}},
  \bibinfo {author} {\bibfnamefont {L.}~\bibnamefont {Jun-Qing}}, \ and\
  \bibinfo {author} {\bibfnamefont {W.}~\bibnamefont {Scheid}},\ }\href
  {\doibase 10.1088/1674-1137/33/S1/028} {\bibfield  {journal} {\bibinfo
  {journal} {Chin. Phys. C}\ }\textbf {\bibinfo {volume} {33}},\ \bibinfo
  {pages} {86} (\bibinfo {year} {2009})}\BibitemShut {NoStop}%
\bibitem [{\citenamefont {Li}\ \emph {et~al.}(2006)\citenamefont {Li},
  \citenamefont {Wang}, \citenamefont {Jia}, \citenamefont {Xu}, \citenamefont
  {Zuo}, \citenamefont {Li}, \citenamefont {Zhao}, \citenamefont {Li},\ and\
  \citenamefont {Scheid}}]{Li_2006}%
  \BibitemOpen
  \bibfield  {author} {\bibinfo {author} {\bibfnamefont {W.}~\bibnamefont
  {Li}}, \bibinfo {author} {\bibfnamefont {N.}~\bibnamefont {Wang}}, \bibinfo
  {author} {\bibfnamefont {F.}~\bibnamefont {Jia}}, \bibinfo {author}
  {\bibfnamefont {H.}~\bibnamefont {Xu}}, \bibinfo {author} {\bibfnamefont
  {W.}~\bibnamefont {Zuo}}, \bibinfo {author} {\bibfnamefont {Q.}~\bibnamefont
  {Li}}, \bibinfo {author} {\bibfnamefont {E.}~\bibnamefont {Zhao}}, \bibinfo
  {author} {\bibfnamefont {J.}~\bibnamefont {Li}}, \ and\ \bibinfo {author}
  {\bibfnamefont {W.}~\bibnamefont {Scheid}},\ }\href {\doibase
  10.1088/0954-3899/32/8/006} {\bibfield  {journal} {\bibinfo  {journal} {J.
  Phys. G}\ }\textbf {\bibinfo {volume} {32}},\ \bibinfo {pages} {1143}
  (\bibinfo {year} {2006})}\BibitemShut {NoStop}%
\bibitem [{\citenamefont {Niu}\ \emph {et~al.}(2021{\natexlab{b}})\citenamefont
  {Niu}, \citenamefont {Chen},\ and\ \citenamefont {Feng}}]{Niu2021}%
  \BibitemOpen
  \bibfield  {author} {\bibinfo {author} {\bibfnamefont {F.}~\bibnamefont
  {Niu}}, \bibinfo {author} {\bibfnamefont {P.-H.}\ \bibnamefont {Chen}}, \
  and\ \bibinfo {author} {\bibfnamefont {Z.-Q.}\ \bibnamefont {Feng}},\ }\href
  {\doibase 10.1007/s41365-021-00946-3} {\bibfield  {journal} {\bibinfo
  {journal} {Nuclear Science and Techniques}\ }\textbf {\bibinfo {volume}
  {32}},\ \bibinfo {pages} {103} (\bibinfo {year}
  {2021}{\natexlab{b}})}\BibitemShut {NoStop}%
\bibitem [{\citenamefont {Gan}\ \emph {et~al.}(2022)\citenamefont {Gan},
  \citenamefont {Huang}, \citenamefont {Zhang}, \citenamefont {Zhou},\ and\
  \citenamefont {Xu}}]{arti22gan}%
  \BibitemOpen
  \bibfield  {author} {\bibinfo {author} {\bibfnamefont {Z.~G.}\ \bibnamefont
  {Gan}}, \bibinfo {author} {\bibfnamefont {W.~X.}\ \bibnamefont {Huang}},
  \bibinfo {author} {\bibfnamefont {Z.~Y.}\ \bibnamefont {Zhang}}, \bibinfo
  {author} {\bibfnamefont {X.~H.}\ \bibnamefont {Zhou}}, \ and\ \bibinfo
  {author} {\bibfnamefont {H.~S.}\ \bibnamefont {Xu}},\ }\href {\doibase
  10.1140/epja/s10050-022-00811-w} {\bibfield  {journal} {\bibinfo  {journal}
  {Eur. Phys. J. A}\ }\textbf {\bibinfo {volume} {58}},\ \bibinfo {pages} {158}
  (\bibinfo {year} {2022})}\BibitemShut {NoStop}%
\bibitem [{\citenamefont {Kajino}(2023)}]{Kajino2023}%
  \BibitemOpen
  \bibfield  {author} {\bibinfo {author} {\bibfnamefont {T.}~\bibnamefont
  {Kajino}},\ }\href {\doibase 10.1007/s41365-023-01196-1} {\bibfield
  {journal} {\bibinfo  {journal} {Nuclear Science and Techniques}\ }\textbf
  {\bibinfo {volume} {34}},\ \bibinfo {pages} {42} (\bibinfo {year}
  {2023})}\BibitemShut {NoStop}%
\bibitem [{\citenamefont {Ming}\ \emph {et~al.}(2022)\citenamefont {Ming},
  \citenamefont {Zhang}, \citenamefont {Xu}, \citenamefont {Sun}, \citenamefont
  {Tian},\ and\ \citenamefont {Ge}}]{Ming2022}%
  \BibitemOpen
  \bibfield  {author} {\bibinfo {author} {\bibfnamefont {X.-C.}\ \bibnamefont
  {Ming}}, \bibinfo {author} {\bibfnamefont {H.-F.}\ \bibnamefont {Zhang}},
  \bibinfo {author} {\bibfnamefont {R.-R.}\ \bibnamefont {Xu}}, \bibinfo
  {author} {\bibfnamefont {X.-D.}\ \bibnamefont {Sun}}, \bibinfo {author}
  {\bibfnamefont {Y.}~\bibnamefont {Tian}}, \ and\ \bibinfo {author}
  {\bibfnamefont {Z.-G.}\ \bibnamefont {Ge}},\ }\href {\doibase
  10.1007/s41365-022-01031-z} {\bibfield  {journal} {\bibinfo  {journal}
  {Nuclear Science and Techniques}\ }\textbf {\bibinfo {volume} {33}},\
  \bibinfo {pages} {48} (\bibinfo {year} {2022})}\BibitemShut {NoStop}%
\bibitem [{\citenamefont {Weizs{\"a}cker}(1935)}]{Weizsacker1935}%
  \BibitemOpen
  \bibfield  {author} {\bibinfo {author} {\bibfnamefont {C.~F.~v.}\
  \bibnamefont {Weizs{\"a}cker}},\ }\href {\doibase 10.1007/BF01337700}
  {\bibfield  {journal} {\bibinfo  {journal} {Zeitschrift f{\"u}r Physik}\
  }\textbf {\bibinfo {volume} {96}},\ \bibinfo {pages} {431} (\bibinfo {year}
  {1935})}\BibitemShut {NoStop}%
\bibitem [{\citenamefont {Strutinsky}(1967)}]{STRUTINSKY1967420}%
  \BibitemOpen
  \bibfield  {author} {\bibinfo {author} {\bibfnamefont {V.}~\bibnamefont
  {Strutinsky}},\ }\href {\doibase
  https://doi.org/10.1016/0375-9474(67)90510-6} {\bibfield  {journal} {\bibinfo
   {journal} {Nuclear Physics A}\ }\textbf {\bibinfo {volume} {95}},\ \bibinfo
  {pages} {420} (\bibinfo {year} {1967})}\BibitemShut {NoStop}%
\bibitem [{\citenamefont {Strutinsky}(1968)}]{STRUTINSKY19681}%
  \BibitemOpen
  \bibfield  {author} {\bibinfo {author} {\bibfnamefont {V.}~\bibnamefont
  {Strutinsky}},\ }\href {\doibase
  https://doi.org/10.1016/0375-9474(68)90699-4} {\bibfield  {journal} {\bibinfo
   {journal} {Nuclear Physics A}\ }\textbf {\bibinfo {volume} {122}},\ \bibinfo
  {pages} {1} (\bibinfo {year} {1968})}\BibitemShut {NoStop}%
\bibitem [{\citenamefont {Oganessian}\ \emph
  {et~al.}(2004{\natexlab{c}})\citenamefont {Oganessian}, \citenamefont
  {Utyonkov}, \citenamefont {Lobanov}, \citenamefont {Abdullin}, \citenamefont
  {Polyakov}, \citenamefont {Shirokovsky}, \citenamefont {Tsyganov},
  \citenamefont {Gulbekian}, \citenamefont {Bogomolov}, \citenamefont {Gikal},
  \citenamefont {Mezentsev}, \citenamefont {Iliev}, \citenamefont {Subbotin},
  \citenamefont {Sukhov}, \citenamefont {Voinov}, \citenamefont {Buklanov},
  \citenamefont {Subotic}, \citenamefont {Zagrebaev}, \citenamefont {Itkis},
  \citenamefont {Patin}, \citenamefont {Moody}, \citenamefont {Wild},
  \citenamefont {Stoyer}, \citenamefont {Stoyer}, \citenamefont {Shaughnessy},
  \citenamefont {Kenneally}, \citenamefont {Wilk}, \citenamefont {Lougheed},
  \citenamefont {Il'kaev},\ and\ \citenamefont
  {Vesnovskii}}]{PhysRevC.70.064609}%
  \BibitemOpen
  \bibfield  {author} {\bibinfo {author} {\bibfnamefont {Y.~T.}\ \bibnamefont
  {Oganessian}}, \bibinfo {author} {\bibfnamefont {V.~K.}\ \bibnamefont
  {Utyonkov}}, \bibinfo {author} {\bibfnamefont {Y.~V.}\ \bibnamefont
  {Lobanov}}, \bibinfo {author} {\bibfnamefont {F.~S.}\ \bibnamefont
  {Abdullin}}, \bibinfo {author} {\bibfnamefont {A.~N.}\ \bibnamefont
  {Polyakov}}, \bibinfo {author} {\bibfnamefont {I.~V.}\ \bibnamefont
  {Shirokovsky}}, \bibinfo {author} {\bibfnamefont {Y.~S.}\ \bibnamefont
  {Tsyganov}}, \bibinfo {author} {\bibfnamefont {G.~G.}\ \bibnamefont
  {Gulbekian}}, \bibinfo {author} {\bibfnamefont {S.~L.}\ \bibnamefont
  {Bogomolov}}, \bibinfo {author} {\bibfnamefont {B.~N.}\ \bibnamefont
  {Gikal}}, \bibinfo {author} {\bibfnamefont {A.~N.}\ \bibnamefont
  {Mezentsev}}, \bibinfo {author} {\bibfnamefont {S.}~\bibnamefont {Iliev}},
  \bibinfo {author} {\bibfnamefont {V.~G.}\ \bibnamefont {Subbotin}}, \bibinfo
  {author} {\bibfnamefont {A.~M.}\ \bibnamefont {Sukhov}}, \bibinfo {author}
  {\bibfnamefont {A.~A.}\ \bibnamefont {Voinov}}, \bibinfo {author}
  {\bibfnamefont {G.~V.}\ \bibnamefont {Buklanov}}, \bibinfo {author}
  {\bibfnamefont {K.}~\bibnamefont {Subotic}}, \bibinfo {author} {\bibfnamefont
  {V.~I.}\ \bibnamefont {Zagrebaev}}, \bibinfo {author} {\bibfnamefont {M.~G.}\
  \bibnamefont {Itkis}}, \bibinfo {author} {\bibfnamefont {J.~B.}\ \bibnamefont
  {Patin}}, \bibinfo {author} {\bibfnamefont {K.~J.}\ \bibnamefont {Moody}},
  \bibinfo {author} {\bibfnamefont {J.~F.}\ \bibnamefont {Wild}}, \bibinfo
  {author} {\bibfnamefont {M.~A.}\ \bibnamefont {Stoyer}}, \bibinfo {author}
  {\bibfnamefont {N.~J.}\ \bibnamefont {Stoyer}}, \bibinfo {author}
  {\bibfnamefont {D.~A.}\ \bibnamefont {Shaughnessy}}, \bibinfo {author}
  {\bibfnamefont {J.~M.}\ \bibnamefont {Kenneally}}, \bibinfo {author}
  {\bibfnamefont {P.~A.}\ \bibnamefont {Wilk}}, \bibinfo {author}
  {\bibfnamefont {R.~W.}\ \bibnamefont {Lougheed}}, \bibinfo {author}
  {\bibfnamefont {R.~I.}\ \bibnamefont {Il'kaev}}, \ and\ \bibinfo {author}
  {\bibfnamefont {S.~P.}\ \bibnamefont {Vesnovskii}},\ }\href {\doibase
  10.1103/PhysRevC.70.064609} {\bibfield  {journal} {\bibinfo  {journal} {Phys.
  Rev. C}\ }\textbf {\bibinfo {volume} {70}},\ \bibinfo {pages} {064609}
  (\bibinfo {year} {2004}{\natexlab{c}})}\BibitemShut {NoStop}%
\bibitem [{\citenamefont {Kaji}\ \emph {et~al.}(2017)\citenamefont {Kaji},
  \citenamefont {Morimoto}, \citenamefont {Haba}, \citenamefont {Wakabayashi},
  \citenamefont {Takeyama}, \citenamefont {Yamaki}, \citenamefont {Komori},
  \citenamefont {Yanou}, \citenamefont {Goto},\ and\ \citenamefont
  {Morita}}]{doi:10.7566/JPSJ.86.085001}%
  \BibitemOpen
  \bibfield  {author} {\bibinfo {author} {\bibfnamefont {D.}~\bibnamefont
  {Kaji}}, \bibinfo {author} {\bibfnamefont {K.}~\bibnamefont {Morimoto}},
  \bibinfo {author} {\bibfnamefont {H.}~\bibnamefont {Haba}}, \bibinfo {author}
  {\bibfnamefont {Y.}~\bibnamefont {Wakabayashi}}, \bibinfo {author}
  {\bibfnamefont {M.}~\bibnamefont {Takeyama}}, \bibinfo {author}
  {\bibfnamefont {S.}~\bibnamefont {Yamaki}}, \bibinfo {author} {\bibfnamefont
  {Y.}~\bibnamefont {Komori}}, \bibinfo {author} {\bibfnamefont
  {S.}~\bibnamefont {Yanou}}, \bibinfo {author} {\bibfnamefont {S.-i.}\
  \bibnamefont {Goto}}, \ and\ \bibinfo {author} {\bibfnamefont
  {K.}~\bibnamefont {Morita}},\ }\href {\doibase 10.7566/JPSJ.86.085001}
  {\bibfield  {journal} {\bibinfo  {journal} {Journal of the Physical Society
  of Japan}\ }\textbf {\bibinfo {volume} {86}},\ \bibinfo {pages} {085001}
  (\bibinfo {year} {2017})},\ \Eprint
  {http://arxiv.org/abs/https://doi.org/10.7566/JPSJ.86.085001}
  {https://doi.org/10.7566/JPSJ.86.085001} \BibitemShut {NoStop}%
\bibitem [{\citenamefont {Oganessian}\ \emph {et~al.}(2005)\citenamefont
  {Oganessian}, \citenamefont {Utyonkov}, \citenamefont {Dmitriev},
  \citenamefont {Lobanov}, \citenamefont {Itkis}, \citenamefont {Polyakov},
  \citenamefont {Tsyganov}, \citenamefont {Mezentsev}, \citenamefont {Yeremin},
  \citenamefont {Voinov}, \citenamefont {Sokol}, \citenamefont {Gulbekian},
  \citenamefont {Bogomolov}, \citenamefont {Iliev}, \citenamefont {Subbotin},
  \citenamefont {Sukhov}, \citenamefont {Buklanov}, \citenamefont {Shishkin},
  \citenamefont {Chepygin}, \citenamefont {Vostokin}, \citenamefont {Aksenov},
  \citenamefont {Hussonnois}, \citenamefont {Subotic}, \citenamefont
  {Zagrebaev}, \citenamefont {Moody}, \citenamefont {Patin}, \citenamefont
  {Wild}, \citenamefont {Stoyer}, \citenamefont {Stoyer}, \citenamefont
  {Shaughnessy}, \citenamefont {Kenneally}, \citenamefont {Wilk}, \citenamefont
  {Lougheed}, \citenamefont {G\"aggeler}, \citenamefont {Schumann},
  \citenamefont {Bruchertseifer},\ and\ \citenamefont
  {Eichler}}]{PhysRevC.72.034611}%
  \BibitemOpen
  \bibfield  {author} {\bibinfo {author} {\bibfnamefont {Y.~T.}\ \bibnamefont
  {Oganessian}}, \bibinfo {author} {\bibfnamefont {V.~K.}\ \bibnamefont
  {Utyonkov}}, \bibinfo {author} {\bibfnamefont {S.~N.}\ \bibnamefont
  {Dmitriev}}, \bibinfo {author} {\bibfnamefont {Y.~V.}\ \bibnamefont
  {Lobanov}}, \bibinfo {author} {\bibfnamefont {M.~G.}\ \bibnamefont {Itkis}},
  \bibinfo {author} {\bibfnamefont {A.~N.}\ \bibnamefont {Polyakov}}, \bibinfo
  {author} {\bibfnamefont {Y.~S.}\ \bibnamefont {Tsyganov}}, \bibinfo {author}
  {\bibfnamefont {A.~N.}\ \bibnamefont {Mezentsev}}, \bibinfo {author}
  {\bibfnamefont {A.~V.}\ \bibnamefont {Yeremin}}, \bibinfo {author}
  {\bibfnamefont {A.~A.}\ \bibnamefont {Voinov}}, \bibinfo {author}
  {\bibfnamefont {E.~A.}\ \bibnamefont {Sokol}}, \bibinfo {author}
  {\bibfnamefont {G.~G.}\ \bibnamefont {Gulbekian}}, \bibinfo {author}
  {\bibfnamefont {S.~L.}\ \bibnamefont {Bogomolov}}, \bibinfo {author}
  {\bibfnamefont {S.}~\bibnamefont {Iliev}}, \bibinfo {author} {\bibfnamefont
  {V.~G.}\ \bibnamefont {Subbotin}}, \bibinfo {author} {\bibfnamefont {A.~M.}\
  \bibnamefont {Sukhov}}, \bibinfo {author} {\bibfnamefont {G.~V.}\
  \bibnamefont {Buklanov}}, \bibinfo {author} {\bibfnamefont {S.~V.}\
  \bibnamefont {Shishkin}}, \bibinfo {author} {\bibfnamefont {V.~I.}\
  \bibnamefont {Chepygin}}, \bibinfo {author} {\bibfnamefont {G.~K.}\
  \bibnamefont {Vostokin}}, \bibinfo {author} {\bibfnamefont {N.~V.}\
  \bibnamefont {Aksenov}}, \bibinfo {author} {\bibfnamefont {M.}~\bibnamefont
  {Hussonnois}}, \bibinfo {author} {\bibfnamefont {K.}~\bibnamefont {Subotic}},
  \bibinfo {author} {\bibfnamefont {V.~I.}\ \bibnamefont {Zagrebaev}}, \bibinfo
  {author} {\bibfnamefont {K.~J.}\ \bibnamefont {Moody}}, \bibinfo {author}
  {\bibfnamefont {J.~B.}\ \bibnamefont {Patin}}, \bibinfo {author}
  {\bibfnamefont {J.~F.}\ \bibnamefont {Wild}}, \bibinfo {author}
  {\bibfnamefont {M.~A.}\ \bibnamefont {Stoyer}}, \bibinfo {author}
  {\bibfnamefont {N.~J.}\ \bibnamefont {Stoyer}}, \bibinfo {author}
  {\bibfnamefont {D.~A.}\ \bibnamefont {Shaughnessy}}, \bibinfo {author}
  {\bibfnamefont {J.~M.}\ \bibnamefont {Kenneally}}, \bibinfo {author}
  {\bibfnamefont {P.~A.}\ \bibnamefont {Wilk}}, \bibinfo {author}
  {\bibfnamefont {R.~W.}\ \bibnamefont {Lougheed}}, \bibinfo {author}
  {\bibfnamefont {H.~W.}\ \bibnamefont {G\"aggeler}}, \bibinfo {author}
  {\bibfnamefont {D.}~\bibnamefont {Schumann}}, \bibinfo {author}
  {\bibfnamefont {H.}~\bibnamefont {Bruchertseifer}}, \ and\ \bibinfo {author}
  {\bibfnamefont {R.}~\bibnamefont {Eichler}},\ }\href {\doibase
  10.1103/PhysRevC.72.034611} {\bibfield  {journal} {\bibinfo  {journal} {Phys.
  Rev. C}\ }\textbf {\bibinfo {volume} {72}},\ \bibinfo {pages} {034611}
  (\bibinfo {year} {2005})}\BibitemShut {NoStop}%
\bibitem [{\citenamefont {Oganessian}\ \emph {et~al.}(2022)\citenamefont
  {Oganessian}, \citenamefont {Utyonkov}, \citenamefont {Ibadullayev},
  \citenamefont {Abdullin}, \citenamefont {Dmitriev}, \citenamefont {Itkis},
  \citenamefont {Karpov}, \citenamefont {Kovrizhnykh}, \citenamefont
  {Kuznetsov}, \citenamefont {Petrushkin}, \citenamefont {Podshibiakin},
  \citenamefont {Polyakov}, \citenamefont {Popeko}, \citenamefont {Sagaidak},
  \citenamefont {Schlattauer}, \citenamefont {Shubin}, \citenamefont
  {Shumeiko}, \citenamefont {Solovyev}, \citenamefont {Tsyganov}, \citenamefont
  {Voinov}, \citenamefont {Subbotin}, \citenamefont {Bodrov}, \citenamefont
  {Sabel'nikov}, \citenamefont {Lindner}, \citenamefont {Rykaczewski},
  \citenamefont {King}, \citenamefont {Roberto}, \citenamefont {Brewer},
  \citenamefont {Grzywacz}, \citenamefont {Gan}, \citenamefont {Zhang},
  \citenamefont {Huang},\ and\ \citenamefont {Yang}}]{PhysRevC.106.024612}%
  \BibitemOpen
  \bibfield  {author} {\bibinfo {author} {\bibfnamefont {Y.~T.}\ \bibnamefont
  {Oganessian}}, \bibinfo {author} {\bibfnamefont {V.~K.}\ \bibnamefont
  {Utyonkov}}, \bibinfo {author} {\bibfnamefont {D.}~\bibnamefont
  {Ibadullayev}}, \bibinfo {author} {\bibfnamefont {F.~S.}\ \bibnamefont
  {Abdullin}}, \bibinfo {author} {\bibfnamefont {S.~N.}\ \bibnamefont
  {Dmitriev}}, \bibinfo {author} {\bibfnamefont {M.~G.}\ \bibnamefont {Itkis}},
  \bibinfo {author} {\bibfnamefont {A.~V.}\ \bibnamefont {Karpov}}, \bibinfo
  {author} {\bibfnamefont {N.~D.}\ \bibnamefont {Kovrizhnykh}}, \bibinfo
  {author} {\bibfnamefont {D.~A.}\ \bibnamefont {Kuznetsov}}, \bibinfo {author}
  {\bibfnamefont {O.~V.}\ \bibnamefont {Petrushkin}}, \bibinfo {author}
  {\bibfnamefont {A.~V.}\ \bibnamefont {Podshibiakin}}, \bibinfo {author}
  {\bibfnamefont {A.~N.}\ \bibnamefont {Polyakov}}, \bibinfo {author}
  {\bibfnamefont {A.~G.}\ \bibnamefont {Popeko}}, \bibinfo {author}
  {\bibfnamefont {R.~N.}\ \bibnamefont {Sagaidak}}, \bibinfo {author}
  {\bibfnamefont {L.}~\bibnamefont {Schlattauer}}, \bibinfo {author}
  {\bibfnamefont {V.~D.}\ \bibnamefont {Shubin}}, \bibinfo {author}
  {\bibfnamefont {M.~V.}\ \bibnamefont {Shumeiko}}, \bibinfo {author}
  {\bibfnamefont {D.~I.}\ \bibnamefont {Solovyev}}, \bibinfo {author}
  {\bibfnamefont {Y.~S.}\ \bibnamefont {Tsyganov}}, \bibinfo {author}
  {\bibfnamefont {A.~A.}\ \bibnamefont {Voinov}}, \bibinfo {author}
  {\bibfnamefont {V.~G.}\ \bibnamefont {Subbotin}}, \bibinfo {author}
  {\bibfnamefont {A.~Y.}\ \bibnamefont {Bodrov}}, \bibinfo {author}
  {\bibfnamefont {A.~V.}\ \bibnamefont {Sabel'nikov}}, \bibinfo {author}
  {\bibfnamefont {A.}~\bibnamefont {Lindner}}, \bibinfo {author} {\bibfnamefont
  {K.~P.}\ \bibnamefont {Rykaczewski}}, \bibinfo {author} {\bibfnamefont
  {T.~T.}\ \bibnamefont {King}}, \bibinfo {author} {\bibfnamefont {J.~B.}\
  \bibnamefont {Roberto}}, \bibinfo {author} {\bibfnamefont {N.~T.}\
  \bibnamefont {Brewer}}, \bibinfo {author} {\bibfnamefont {R.~K.}\
  \bibnamefont {Grzywacz}}, \bibinfo {author} {\bibfnamefont {Z.~G.}\
  \bibnamefont {Gan}}, \bibinfo {author} {\bibfnamefont {Z.~Y.}\ \bibnamefont
  {Zhang}}, \bibinfo {author} {\bibfnamefont {M.~H.}\ \bibnamefont {Huang}}, \
  and\ \bibinfo {author} {\bibfnamefont {H.~B.}\ \bibnamefont {Yang}},\ }\href
  {\doibase 10.1103/PhysRevC.106.024612} {\bibfield  {journal} {\bibinfo
  {journal} {Phys. Rev. C}\ }\textbf {\bibinfo {volume} {106}},\ \bibinfo
  {pages} {024612} (\bibinfo {year} {2022})}\BibitemShut {NoStop}%
\end{thebibliography}
%

\end{document}